\documentclass[runningheads]{llncs}
\usepackage[T2A,OT2,OT4,T1]{fontenc}
\usepackage[utf8]{inputenc}
\usepackage[polish,greek,russian,english]{babel}

\RequirePackage{amsfonts}
\usepackage{comment}
\usepackage{filecontents}
\usepackage[numbers,sort&compress]{natbib}
\usepackage{amsmath}
\usepackage[mathscr]{eucal}
\usepackage{bbm}
\usepackage{dsfont}
\usepackage{mathtools}
\usepackage{wrapfig,booktabs}
\usepackage{graphicx}
\graphicspath{{./Figures/},{../Figures/},{./Pictures/}}
\usepackage{pgf}
\usepackage{tikz}\usetikzlibrary{arrows,automata}
\usepackage{float}
\usepackage{enumerate}
\usepackage{hyperref}
\hypersetup{
     pdftitle={Operation comfort and the importance of system components}
     pdfauthor=Krzysztof Szajowski and Małgorzata Średnicka, 
     colorlinks,
     urlcolor=blue,
     filecolor=magenta,
     citecolor=green, 
     linkbordercolor={1 1 1}, 
     citebordercolor={1 1 1},  
     urlbordercolor={ 1 1 1}  
} 
\RequirePackage[hyperpageref]{backref} 
    \renewcommand*{\backref}[1]{}  
    \renewcommand*{\backrefalt}[4]{
       \ifcase #1 
          No cited.
       \or
          Cited on p. #2.
       \else
          Cited on pp. #2.
       \fi}

\newcommand{\authorcontributions}[1]{%
\vspace{6pt}\noindent{\fontsize{9}{9}\selectfont\textbf{Author Contributions:} {#1}\par}}

\newcommand{\funding}[1]{
\vspace{6pt}\noindent{\fontsize{9}{9}\selectfont\textbf{Funding:} {#1}\par}}

\newcommand{\conflictsofinterest}[1]{%
\vspace{6pt}\noindent{\fontsize{9}{9}\selectfont\textbf{Conflicts of Interest:} {#1}\par}}

\newcommand{\abbreviations}[1]{%
\vspace{12pt}\noindent{\selectfont\textbf{Abbreviations}\par\vspace{6pt}\noindent {\fontsize{9}{9}\selectfont #1}\par}}

\usepackage{glossaries}

\makeglossaries

\newglossaryentry{$Q(t)$}{
name=Survival Function, 
description={It is the reliability function (alternative names: the survivor function): $Q(t)=\bar{F}(t)=1-F(t)$}
}
\newglossaryentry{$phi$}{
name=Structure Function, 
description={It is the structure function: $phi$}
}
\def\subclass#1{\par\addvspace\medskipamount{\rightskip=0pt plus1cm
\def\and{\ifhmode\unskip\nobreak\fi\ $\cdot$
}\noindent\subclassname\ignorespaces #1 \par}}
\DeclareMathSymbol{\preccurlyeq}  {\mathrel}{AMSa}{"34}
\newtheorem{fact}[theorem]{Fact}
\newcommand{\orcid}[1]{\href{https://orcid.org/#1}{\includegraphics[scale=.05]{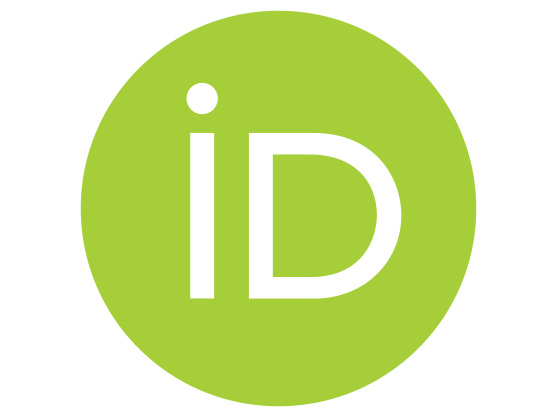}}}
\newcommand*{\doi}[1]{\href{http://dx.doi.org/#1}{doi: #1}}
\newcommand*{\MR}[1]{\href{http://www.ams.org/mathscinet-getitem?mr=#1&return=pdf}{MR #1}}
\newcommand*{\ZBL}[1]{\href{http://www.zentralblatt-math.org/zmath/en/advanced/?q=an:#1&format=complete}{Zbl #1}}
\newcommand\myatop[2]{\genfrac{}{}{0pt}{}{#1}{#2}}
\def\vecx{\overrightarrow{x}}
\def\vecy{\overrightarrow{y}}
\def\vecf{\overrightarrow{\textbf{f}}}
\def\vecp{\overrightarrow{\textbf{p}}}
\def\vecq{\overrightarrow{\textbf{q}}}
\def\vecv{\overrightarrow{\textbf{v}}}
\def\vecI{\overrightarrow{\textbf{I}}}
\def\vecX{\overrightarrow{X}}
\def\vecF{\overrightarrow{F}}

\def\vecdelta{\overrightarrow{\delta}}

\def\bbC{\mathbb{C}}
\def\bbE{\mathbb{E}}
\def\vecQ{\overrightarrow{Q}}
\def\One{\mathbb{I}}
\def\one{\mathbb{I}}
\def\bbN{\mathbb{N}}

\def\bD{\textbf{D}}
\def\bE{\textbf{E}}
\def\bP{\textbf{P}}

\def\cB{{\mathcal B}} 
\def\cF{{\mathcal F}}
\def\scrS{{\mathscr S}}

\def\fri{\mathfrak i}

\begin{document}
\title{Operation comfort vs. the importance of system components
}
%
\author{Krzysztof J. Szajowski\inst{1}\orcid{ 0000-0001-9834-9929} \and
Małgorzata Średnicka\inst{2}\orcid{1111-2222-3333-4444} 
}
\titlerunning{Operation comfort vs. the importance measures}
\authorrunning{K.~Szajowski and M. Średnicka}
%
\institute{Wrocław University of Science and Technology, Faculty of Pure and Applied Mathematics, Wybrzeże Wyspiańskiego 27, 50-370 Wrocław, Poland
\email{Krzysztof.Szajowski@pwr.edu.pl}\\
\url{http://szajowski.wordpress.com/} \and
Wrocław University of Science and Technology\\
\email{margaret.srednicka@gmail.com}
}
\maketitle              
\begin{abstract}
The paper focuses on portraying importance measures that are reasonably helpful in analyzing system reliability and its development. The presented measures concern coherent binary and multistate systems and help to distinguish the most influential elements of the system, which require more attention. Importance measures are presented for systems with known structure (e.g. parallel, series) and for repairable or nonrepairable components


\subclass{ MSC  90B25 \and (62N05; 60K10) }

\end{abstract}
\centerline{\large Table of Contents}
\contentsline {section}{\numberline {1}Introduction}{2}{section.1.1}%

\contentsline {subsection}{\numberline {1.1}Preliminaries}{2}{subsection.1.1.1}%
\contentsline {subsection}{\numberline {1.2}Availability for planned tasks determines reliability.}{4}{subsection.1.1.2}%
\contentsline {subsection}{\numberline {1.3}Raised the role of the element in failure.}{5}{subsection.1.1.3}%
\contentsline {subsection}{\numberline {1.4}General systems classification}{6}{subsection.1.1.4}%
\contentsline {subsection}{\numberline {1.5}Review of importance measure concepts}{7}{subsection.1.1.5}%
\contentsline {subsection}{\numberline {1.6}Cooperative games vs. semi-coherent systems}{8}{subsection.1.1.6}%
\noindent
\contentsline {section}{\numberline {2}Binary systems}{8}{section.1.2}%

\contentsline {subsection}{\numberline {2.1}Preliminary remarks}{8}{subsection.1.2.1}%
\contentsline {subsection}{\numberline {2.2}Coherence and system structure}{9}{subsection.1.2.2}%
\contentsline {subsection}{\numberline {2.3}Reliability importance.}{10}{subsection.1.2.3}%
\contentsline {subsection}{\numberline {2.4}Lifetime importance measure.}{12}{subsection.1.2.4}%
\contentsline {subsection}{\numberline {2.5}Module importance}{14}{subsection.1.2.5}%
\contentsline {subsection}{\numberline {2.6}Structural importance}{17}{subsection.1.2.6}%
\contentsline {subsection}{\numberline {2.7}Importance measure based on multilateral stopping problem.}{23}{subsection.1.2.7}%
\contentsline {section}{\numberline {3}Concluding remarks}{26}{section.1.3}%
\contentsline {subsection}{\numberline {3.1}Summary}{26}{subsection.1.3.1}%
\contentsline {subsection}{\numberline {3.2}Exploratory importance measure research.}{26}{subsection.1.3.2}%
\contentsline {subsection}{\numberline {3.3}Important direction of further investigations.}{27}{subsection.1.3.3}%
\contentsline {section}{\numberline {}Appendices. }{27}{section.A}%
\contentsline {subsection}{\numberline {A}Structure functions. }{27}{section.A.1}%
\contentsline {subsection}{\numberline {B}The simple game}{29}{section.A.2}%
\contentsline {subsection}{\numberline {C}Power indexes}{29}{section.A.3}%
\contentsline {section}{\numberline {}Bibliography. }{30}{section.A}%

\section{Introduction}
\subsection{Preliminaries}
Let's consider a system\footnote{System (in Ancient Greek:  {\selectlanguage{greek}σύστημα }--romanized:~systema -- a~complex thing) -- a set of interrelated elements realizing the assumed goals as a whole.}, a complex structure with specific functionality. Contemporary systems are characterized by their structural complexity. In the process of designing the system, the most important thing is its preparation for the implementation of the assumed goals. The mathematical model of the system is based on the set theory as the family of subsets of given set $\bbC=\{c_1,\ldots,c_n\}$ having some properties. An example is technical devices whose design is dictated by the need to perform specific functions. The constructed system should function in a planned and predictable manner. This property is a requirement that should also be considered in the design and construction (fabrication) process. The goal is therefore to reduce the risk\footnote{It is difficult to define \emph{risk} in general. In short, when we think about risk, we mean the possibility of an unexpected loss caused by an unpredictable event or harmful behavior (human, machine, animal, nature). One can think about possibility of loss or injury. From the other side, the risk is the chance or probability that a person (a system) will be harmed or experience an adverse health (functioning) effect if exposed to a hazard. It may also apply to situations with property or equipment loss, or harmful effects on the environment. Therefore, we are talking about reducing ownership and loss as a result of a random event. Risk reduction means minimizing the chance of a loss occurring or limiting its size. In order to better understand the risk and possibilities of risk management, the task of measuring risk has been set. The task is not formulated so that its solution is universal. This allowed to determine the desired properties of such measures \cite{ArtDelEbeHea1999:Coherent}. } of a break in the planned operation of the system. Therefore, ensuring reliability and proper operation is of great importance in system analysis and management of its operation. One of the measures to assess the quality of a solution is system performance. Correct and expected operation gives the expected results - both in terms of size, time of achievement and costs (outlays) of receiving them. These expectations are achieved by ensuring reliable system operation. The performance of the system is therefore affected by the reliability of its components and its structure. At the same time, not only the reliability of the system depends on these factors. In the event of a failure, it is important to be able to \emph{localize the damage} more easily and to remove it (repair it). Therefore, it is obvious that not all elements have the same effect on the functioning of the system. To improve system reliability and readiness, as well as streamline maintenance and repair activities, the importance of system components should be explored for both reliability and maintenance - including diagnostics, maintenance and repairs. Without proper analysis, it is impossible to predict the significance of individual elements for these features. Individual elements may affect each of them to a different degree. There are known results on the evaluation of the weight of components on the reliability of the system. The introduced measures of significance of elements on reliability will be the basis for the introduction of diagnostic algorithms, about the possibility of which they wrote at the end of his seminal paper by \citeauthor{Bir1969:Multivariate}(\citeyear{Bir1968:importance,Bir1969:Multivariate}) (v. \citeauthor{BarFusSin1975:DedBirnbaum}~(\citeyear{BarFusSin1975:DedBirnbaum})). The indication of these algorithms is the subject of this study. 

In order to determine the significance of the reliability of individual system components to the reliability of the whole system, measures are constructed that are sensitive to modifications in the system. This allows the rationalization of the structure and subsequent planning for optimal maintenance. The issues are complex due to the fact that it is necessary to take into account both the effective reliability of the constructed system and the cost of maintaining it in readiness in a given period. Profitability analysis is of great importance. It is natural to formulate the problem by defining the primary goal of minimizing the cost while guaranteeing the expected levels of reliability. With this approach, it is possible to define weights for the cost of individual elements in a given time horizon, while ensuring a certain level of security or readiness. This approach can be found at paper by \citeauthor{WuCoo2013:Cost-based}(\citeyear{WuCoo2013:Cost-based}). At the same time, one should not forget about the other key goals and parameters in system analysis. Their inclusion in the balanced model is possible with the use of natural methods of analysis in the formulation of many criteria, based on elements of game theory.

We are trying to present the issue comprehensively, although there is currently no consistent approach on the way of determining the importance of elements in the system. This is because the loss of functionality of an element often does not clearly affect the system's ability to perform tasks. This aspect is highlighted by numerous examples presented in the literature, which show a significant impact of the state of the environment in which the systems are operated (time of day, weather conditions, environmental pollution). In addition, attention should be paid to the cause-effect relationships of the work of the elements. We often deal with a sequential progression of damage and degradation of elements, which means that it is possible to propose a modeling method without the possibility of creating a universal model, the calibration of which allows for a proper description of the analyzed problem. The methodological limitations mentioned here mean that the proposed methods are a development of the problems that we mention, but we do not exclude that the approach may also enrich other analyzes based on other premises and conditions. In order to organize the methodology, we will use the systems classification, which will allow us to formulate assumptions. Wherever possible, we provide the source of inspiration (a description of an issue in which the proposed approach can be modeled, or we cite sources in the literature that use an analogous model), although we realize that getting to the original formulation of a concept or approach does not have to be the best justification and motivation for the proposed approach. 

\subsection{Availability for planned tasks determines reliability.} We analyze the system (layout, structure) as one whole, carrying out a specific simple task. We consider systems that are made up of elements. The system is operational if it can accomplish the task for which it was created. With this formulation, we assume that the task execution time is infinitely short, so the possibility of failure in the course of the task can be neglected. The analysis of the role of the components in such a system comes down to the assessment of the impact of the reliability of a specific element on the reliability of the whole. For this category of tests, measures of the importance of elements will be helpful, which allow for the assessment (measure) of the improvement in the system reliability resulting from the improvement of the reliability of a given component. Such measures are useful in determining the components whose reliability should be improved to obtain the maximum improvement in system reliability. Examples of such measures can be found in the works \citeauthor{BirnbaumZ.W.1961MSaS}~(\citeyear{BirnbaumZ.W.1961MSaS}), \citeauthor{Bir1969:Multivariate}~(\citeyear{Bir1969:Multivariate}), \citeauthor{Nat1985:New}~(\citeyear{Nat1985:New}), \citeauthor{BolEl-NPro1988:Active}~(\citeyear{BolEl-NPro1988:Active}).

We want, at this level of generality, to measure the weight of an element related to its place in the structure, and structure has a role when the system is intentionally designed. This analysis is also performed when the reliability of the components is unknown. Hence we say that we are looking for a measure of the significance of the structural element (structural importance measure).

The factor that we want to include in the analysis is not only the position, but also the reliability of the element. While still maintaining the assumption that the system takes an infinitely short time to complete a task, we introduce a measure of the element's significance for reliability reasons (reliability importance measure).

If the time needed to perform the task cannot be omitted, or the tasks are repeated, and we know the reliability functions of the elements, the element significance measure should also take into account the changes in the reliability of elements over time. The inclusion of the reliability function in the element significance analysis can be performed in various ways: global, local or for a fixed time period ( various lifetime importance measure).

The aspects presented relate to the readiness to perform the task, excluding the need for maintenance and repair, including the costs of these activities (cost of parts, repair and maintenance time, penalties for non-availability). In system maintenance tasks, in determining component importance, issues such as detecting damaged components at system shutdown are important. The element that should be checked in the first place (because it is most suspected of causing a failure) can be treated as important for the efficient conduct of maintenance or repair (v. e.g. \citeauthor{Pin2004:PhD}~(\citeyear{Pin2004:PhD})). 

\subsection{Raised the role of the element in failure.} At the time of failure (and not analysis during construction), the system analyzes the maintenance team. It can monitor the state of the system. He wants to find out what the elements meant for the observed state. To facilitate this analysis, we determine the posterior weights of the elements. Otherwise, in these considerations, the measure of importance of a component (group of components) in a given system is based on the quantification of the "role" of that component (group of components) in the failure of that system. Examples of such measures can be found in \citeauthor{FusVes1972:Overview}~(\citeyear{FusVes1972:Overview}), \citeauthor{BarPro1975:Importance}~(\citeyear{BarPro1975:Importance}), \citeauthor{El-NProSet1978:simple}~(\citeyear{El-NProSet1978:simple}), \citeauthor{El-NSet1991:modules}~(\citeyear{El-NSet1991:modules}) and \citeauthor{AboEl-nSet1994:modules}~(\citeyear{AboEl-nSet1994:modules}). Defined measures (indices) of significance allow us to identify the components (groups) that are probably responsible for "causing" a system failure. Establishing these indexes, in turn, leads to an effective control and maintenance principle, as well as optimizing the storage of spare parts and optimal allocation of repairs to the appropriate maintenance technicians of the relevant system components.

The purpose of such research is to propose new importance measures for degrading components (v. \citeauthor{CaoLiuFan2019:modules}~(\citeyear{CaoLiuFan2019:modules})). The motivation is based on Shapley value, which can provide answers about how important players are to the whole cooperative game and what payoff each player can reasonably expect. The proposed importance measure characterizes how a specific degrading component contributes to the degradation of system reliability by using Shapley value. Degradation models are also introduced to assess the reliability of degrading components. The reliability of system consisting independent degrading components is obtained by using structure functions, while reliability of system comprising correlated degrading components is evaluated with a multivariate distribution. The ranking of degrading components according to this importance measure depends on the degradation parameters of components, system structure and parameters characterizing the association of components. A reliability degradation of engineering systems and equipment are often attributed to the degradation of a particular or set of components that are characterized by degrading features. This approach reflects the responsibility of each degrading component for the deterioration of system reliability. The results are also able to give timely feedback of the expected contribution of each degrading component to system reliability degradation.   

\subsection{General systems classification}
The systems can be split into two categories:
\begin{enumerate}[(i)]\itemsep1pt \parskip1pt \parsep1pt
    \item Binary systems (BS)\label{Binarysystem}
    \item Multistate systems (MSS) \label{MSS}
\end{enumerate}
A binary system \eqref{Binarysystem} is a system comprised of $n$ elements. It has precisely two states: $0$ - when the system is failed and $1$ - when the system is functioning. However, term "binary" pertains to the components of the system that define it. In this case components may be in only one of two states $1$ - when the component is functioning perfectly and $0$ - when the element is absolutely damaged. Nevertheless, binary systems not always meet the real life problems. Frequently we have to reckon with elements that undergo only partial failure, but do not cease to perform their operation and do not cause the entire system to cease function. This is the case of the multistate systems \eqref{MSS} with the same properties as \eqref{Binarysystem} beside states of components. Binary systems are discussed in chapter \ref{BinarySysMeasure}, while the discussion of multistate systems are moved to next paper. 

There are three main classes of importance measures (v. \citeauthor{Bir1969:Multivariate}(\citeyear{Bir1969:Multivariate}), \citeauthor{AmrKam2017:OverviewIM}(\citeyear{AmrKam2017:OverviewIM}))
\begin{enumerate}[(i)]\itemsep1pt \parskip1pt \parsep1pt
    \item Reliability importance measure \label{intro:reliab}
    \item Structural importance measure \label{intro:struct}
    \item Lifetime importance measure \label{intro:life}
		\item Failure and its recovery costs importance measure \label{intro:recovery}
\end{enumerate}
Reliability importance measure \eqref{intro:reliab} focuses on the change in the reliability of the system due to reliability change of the particular component. The measure is evaluated with respect to the specific finite period of time and depends on the components reliability and on the system structure. Nonetheless, if reliability of the components are unknown, then we consider the case of the structural importance measure \eqref{intro:struct}. To apply it we are obligated to know the structure of the system. Hence, this measure indicates importance of the system by checking significance of the positions occupied by individual components. The lifetime importance measure \eqref{intro:life} depends on the lifetime distribution of component and also on component position in the system. This measure can be divided into two categories with respect to being the function of time: Time Independent Lifetime importance and Time Dependent Lifetime importance. Lastly but not least, the cost of failure and its recovery importance measure \eqref{intro:recovery} depends on the lifetime distribution of component, its position in the system and loss related to non-availability of the system, diagnosis and repair. It is a new look at the importance of the components of a complex system. The analysis and significance measure proposed in this paper is based on the possibility of observing the components and a rational system maintenance policy, which consists in stopping the system for maintenance and repairs at a time when it pays off to a sufficient number of components. The details are based on a cooperative analysis of costs and losses in the operation of such a system (v. the section~\ref{GameImpMeasure}, \citeauthor{szayas95:voting}~(\citeyear{szayas95:voting})).

\subsection{Review of importance measure concepts}
Since \citeauthor{Bir1968:importance}(\citeyear{Bir1968:importance,Bir1969:Multivariate}) the importance measures were investigated and extended in various directions (v. \citeauthor{AmrKam2017:OverviewIM}(\citeyear{AmrKam2017:OverviewIM})). \citeauthor{Ram1990:Simple}(\citeyear{Ram1990:Simple}) shows the relation of these idea to the research on the cooperative games. These relationships can be helpful in determining the importance of elements for the reliability of the system and at the same time a role in the possibility of efficient diagnosis in the event of a failure, as well as in determining the rules of procedure for removing a failure. Removing the failure causes that the features of the element and the repaired module are restored. However, it should be remembered that the method of repair and the quality of the elements used reproduce the original features to varying degrees (v. e.g. \citeauthor{NavArrSua2019:Minimal}(\citeyear{NavArrSua2019:Minimal})). This has an impact on further operation, diagnosis and maintenance (uplift). Rules are easier to set when they are associated with objective measures of the features of components, modules and the system. Analysis of significance measures in the context of repairs helps to understand such relationships. 
Let us therefore establish these relationships (v \citeauthor{DoBer2020:Conditional}~(\citeyear{DoBer2020:Conditional})).  

\begin{definition}[The structure]\label{StructureDef} For a non-empty and finite set $\mathbf N$\footnote{The list of symbols and abbreviations used in the work has been collected in the section abbreviation on page~\pageref{skroty}.}, we denote by $\mathcal P$ the family of subsets $\mathbf N$ having the following properties
\begin{enumerate}[(1)]\itemsep1pt \parskip1pt \parsep1pt
    \item $\emptyset\in {\mathcal P}$;
    \item ${\mathbf N}\in {\mathcal P}$; 
    \item $S\subseteq T\subseteq {\mathbf N}$ and $S\in {\mathcal P}$ imply $T\in {\mathcal P}$. 
\end{enumerate}
The family $\mathcal P$ is called structure.
\end{definition}
This basic structure has been studied in many areas under a variety of names. The monograph by \citeauthor{Ram1990:Simple}(\citeyear{Ram1990:Simple}) unified the definitions and concepts in two main fields of application, that is cooperative game theory (simple games) (v. Appendix~\ref{AppSimpleGame}, Chapt.~10 in \citeauthor{Tij2003:GTIntro}(\citeyear{Tij2003:GTIntro})) and reliability theory (semi-coherent and coherent structures, v. \citeauthor{EsaPro1963:Coherent}~(\citeyear{EsaPro1963:Coherent}), \citeauthor{1978CSwM}~(\citeyear{1978CSwM}), \citeauthor{OhiFumio2010}~(\citeyear{OhiFumio2010})).

In reliability theory, consider the set ${\mathbf N}=\{1,2,\ldots,n\}$ of components with which a system $g$ has been built. The state of the system as well as any component can either be $0$ (a failed state) or $1$ (a functioning state). The knowledge of the system is represented by the knowledge of the structure function of the system which is defined as a switching function (boolean) $g:\{0,1\}^n\rightarrow\{0,1\}$ of $n$ variables (or $n$ dimensional vector $\vec{x}$)\footnote{With the same symbol, we denote the system and the analytical description of the system using the structure function wherever it does not lead to misunderstandings.}. The structure function $g$ (simply the structure $g$) is called semi-coherent if 
\begin{enumerate}
    \item[(1)] $g$ is monotone, i.e. $\vecx\preceq \vecy$ implies $g(\vecx)\leq g(\vecy)$; 
    \item[(2)] $g(\vec0)=0$ and $g(\vec1)=1$.
\end{enumerate} 
The semi-coherent structure can be called coherent when all its elements are significant. A subset $A\subset {\mathbf N}$ is called a path set of $g$, if $g(\vec{1}^A,\vec{0}^{N\setminus A})=1$, i.e. the system is working if the items forming the set $A$ [resp. $N\setminus A$] are working [resp. failed]. Similarly, $A\subset {\mathbf N}$ is called a cut set of $g$, if $g(\vec{0}^A,\vec{1}^{N\setminus A})=0$. Obviously, the assemblage of path [cut] sets of a semi-coherent structure $g$ satisfies the three properties of the basic structure mentioned at the beginning.

\subsection{Cooperative games vs. semi-coherent systems} 
\cite[Sec. 2]{Ram1990:Simple}  indicates the correspondence between the terminology of cooperative game theory and reliability by means of a list of equivalent notions: player or component; simple game or semi-coherent structure; characteristic function or structure function; winning [blocking] coalition or path [cut] set; minimal winning [blocking] coalition or minimal path [cut] set. The review of the various types of simple games and semi-coherent structures encountered in the literature are mentioned there. The most interesting is \cite[Ch. 3]{Ram1990:Simple}, where detailed study of the problem of assessing the importance [power] of components [players] comprising the system [game] is described. The emphasis is on the probabilistic approach to the quantification of relative importance. 

\section{\label{BinarySysMeasure}Binary systems}
\subsection{Preliminary remarks}

Importance measures are helpful in deciding on the development of which element to emphasize in order to improve the functioning of the system, through indicating those more meaningful. A system yield function was a concept of a general measure of importance, firstly introduced by \citeauthor{Bir1968:importance}(\citeyear{Bir1968:importance}). His idea take into account the structure of the system only. Further, the research on the topic went in various direction (cf. \citeauthor{Xie1987:Importance}(\citeyear{Xie1987:Importance})).

New variants of importance measures can be found in \citeauthor{DuiSiWuYam2017:Importance}(\citeyear{DuiSiWuYam2017:Importance}), \citeauthor{WuCoo2013:Cost-based}(\citeyear{WuCoo2013:Cost-based}), \citeauthor{DutRau2015:ExtensionIM}(\citeyear{DutRau2015:ExtensionIM}). Importance measures have been widely used as important decision-aiding indicators in various purposes such as reliability studies, risk analysis and maintenance optimization. A novel time-dependent importance measure for systems composed of multiple non repairable components is proposed by \citeauthor{DoBer2020:Conditional}~(\citeyear{DoBer2020:Conditional}). The proposed importance measure of a component (module of components) is defined as its ability to improve the system reliability during a mission given the current conditions (states or degradation levels) of its components. To take into account economic aspects, like e.g. maintenance costs, economic dependence between components and the cost benefit thanks to maintenance operations, an extension of the proposed importance measure is then investigated. Thanks to these proposed importance measures, the component (group of components) can be \emph{rationally} selected for preventive maintenance regarding to the reliability criteria or the financial issues. The new treatment of the mentioned topic is the subject of the section~\ref{GameImpMeasure}.

\subsection{Coherence and system structure} 
In this paper, we will consider coherent structures, i.e. that are nondecreasing functions. We call these structures monotonic. We will not consider structures whose state does not depend on the states of their elements. 
\begin{definition}\label{def:coherent}
The structure $\phi$ is called semi-coherent if for states $\vecx$ and $\vecy$, such that $\vecx\preceq\vecy$ implies  
\begin{equation*}
       \phi(\vecx) \leq \phi(\vecy), 
\end{equation*}
and coherent if additionally it complies with $\phi(\vec1) = 1$ and $\phi(\vec0) = 0$.
\end{definition}

In multi-component system to classify a structure as coherent, we have to introduce more notation and some properties \cite{BirnbaumZ.W.1961MSaS}, \cite{Bir1969:Multivariate}. Thus, at the very beginning we assume that $n$ components comprise the system, denoted by
$\vec{c} = (c_1, c_2, \dots, c_n)$. Of the two available states (\textbf{S})- failed (\textbf{F}) or functioning (working \textbf{W}) - each component can only have one, what can be defined by a binary indicator variable $x_i=\One_{\textbf{W}}(c_i)$, $c_i\in\{\textbf{F}, \textbf{W}\}$
for every $i=1,2,\ldots,n$. In other words, it is a state vector (vector of component states) $\vecx=(x_1,x_2,\ldots,x_n)$. The comparison of the state vectors can be described with following notation \cite{Bir1969:Multivariate} based on the component states for $i=1,\ldots,n$: 
\begin{eqnarray*}
    & \vecx = \vecy &\mbox{\quad if $y_i = x_i$,} \\
    & \vecx \preceq \vecy &\mbox{\quad if $x_i\geq y_i$,} \\
    & \vecx \prec \vecy &\mbox{\quad if $\vecx \preceq \vecy$ and  $\vecx \neq \vecy$} 
\end{eqnarray*}
Moreover, we assume that a system composed of $n$ elements whose states are binary also has only two states possible - failed or functioning. Let $\phi:\{0,1\}^n\rightarrow\{0,1\}$ be the structure function. If inequality $x_i \leq y_i$ for $i=1,\ldots,n$ fulfills conditions from the definition \ref{def:coherent} and the structure is monotonic and irreducible, then the structure function $\phi$ is called coherent.

The structure function $\phi$ for every $j=1,2,\ldots,n$ may be presented in the manner of 
\begin{align}
    \phi (\vecx)   &= x_j \cdot \delta_j(\vecx) + \mu_j(\vecx) \label{structfun1} \\
\intertext{where}		
     \delta_j(\vecx)&=\phi(1,\vecx_{-j}) - \phi(0,\vecx_{-j})  \label{structfun2} \\
     \mu_j(\vecx)   &=\phi(0,\vecx_{-j}).   \label{structfun3}
\end{align}
Hence, the component $c_j$ with the state $x_j$ does not influence $ \delta_j(\vecx)$ and $\mu_j(\vecx)$.

\subsection{Reliability importance.} If for $i=1,\ldots,n$ we consider independent elements $X_i$, then the system reliability is defined as a function of reliability of its components, which is equal to the probability that the whole system will keep functioning. Let assume the coherent system with a vector $\vecp=(p_1,\dots,p_n)$ of the components reliabilities, and in that case the reliability function is expressed as 
\begin{equation} \label{reliabfun}
    h(\vecp) = \bP\{\omega:\phi(\vecX(\omega))=1|\vecp\} = \bE[\phi(\vecX)|\vecp],
\end{equation}
where $h(\vecp)$ is the reliability of the structure $\phi$ as the function of the reliability of their components. From equations \eqref{structfun1} and \eqref{reliabfun} we have 
\begin{equation} \label{hp}
    h(\vecp) = p_i\cdot \bE[\delta_i(X)] + \bE[\mu_i(X)]
\end{equation}
for every $i=1,\ldots,n$ and from \eqref{structfun1} and \eqref{hp} we obtain the \emph{reliability importance} of the component $c_i$ in the system $\phi$ 
\begin{equation}\label{Gradh}
    I_{\phi}(i;\vecp)=I_h(i;\vecp)\stackrel{\text{\cite{spivak1965calculus}}}{=}\bD_{p_i}h(\vecp)=\frac{\partial }{\partial p_i}h(\vecp) =\frac{\partial h(\vecp)}{\partial p_i}\stackrel{\eqref{hp}}{=} \bE[\delta_i (\vecX)] ,
\end{equation}
which was first introduced by \citeauthor{Bir1969:Multivariate} (\citeyear{Bir1969:Multivariate}).
These \emph{importance measures} are known as a vector $\vec{B}(\vecp)$ having coordinates
\begin{equation}\label{eq:BirnbaumMeasure}
    B(i|\vecp) = \bD_{p_i} h(\vecp)=\bD_{1-p_i}(1-h(\vecp)), \qquad i=1,2,...,n,
\end{equation}
where $B(i|\vecp)$ is $\vecp$ dependent. If reliabilities $\vecp$ are unknown, we obtain the \emph{structural importance}, defined as
\begin{equation}\label{eqBirnbaumStrImp}
    B(i|\vecp) = I_{\phi}(i;\vecp) =\bD_{p_i} h(\vecp)\Bigg\rvert _{p_1=\ldots=p_n=\frac{1}{2}}, \qquad i=1,2,...,n,
\end{equation}
what will be discussed in section \ref{sectionStructuralImportance}. The \emph{reliability importance} (v. \cite{Bir1969:Multivariate}) of a component $c_i$  is defined as 
\begin{align}
\nonumber    I_{\phi}(i,r;\vecp) &= \bP\{\phi(\vecX) = r | X_i =r;\vecp\} - \bP\{\phi(\vecX) = r|\vecp\},\\
\nonumber    										 &= \bP\{\phi(\vecX) = r | (r,\vecp_{-i})\} - \bP\{\phi(\vecX) = r|\vecp\},\\
\intertext{for the functioning of the structure $\phi$ with $r=1$, while for the failure of the structure $\phi$ with $r=0$. Hence, the \emph{compound reliability importance} of the component $c_i$ for the structure $\phi$ is}
\label{eqSumCompRel}    I_{\phi}(i;\vecp) &= I_{\phi}(i,1;\vecp) + I_{\phi}(i,0;\vecp),\\
\intertext{what is exactly equal to}
\label{eqDerivDelta}    I_{\phi}(i;\vecp) &= \frac{\partial h(\vecp)}{\partial p_i} = \bE[\delta_i(\vecX)]
\end{align}
We can easily get \eqref{eqDerivDelta} from \eqref{eqSumCompRel}
\begin{align*}
I_{\phi}(i;\vecp) &= I_{\phi}(i,1;\vecp) + I_{\phi}(i,0;\vecp)\\
									&= \bP\{\phi(\vecX) = 1 | X_i =1;\vecp\} - \bP\{\phi(\vecX) = 1;\vecp\}\\
									&\quad + \bP\{\phi(\vecX) = 0 | X_i =0;\vecp\} - \bP\{\phi(\vecX) = 0;\vecp\}\\
									&=\bP\{\phi(\vecX) = 1 | X_i =1;\vecp\}-(1-\bP\{\phi(\vecX) = 0 | X_i =0;\vecp\})\\
									&=\bP\{\phi(\vecX) = 1 | X_i =1;\vecp\}-\bP\{\phi(\vecX) = 1 | X_i =0;\vecp\}.
\end{align*}
Hence, the equivalent definition of the reliability importance \cite{zaleskafornal} is 
\begin{equation}\label{rownanieReliabImportPrec1}
    I_{h}(i;\vecp) = h(1,\vecp_{-i}) - h(0,\vecp_{-i})= \bE \big[ \phi (1_i,\vecX) - \phi (0_i,\vecX) \big]=\bE \delta_i(\vecX).
\end{equation}
For the coherent system, the reliability of each element and the reliability importance belongs to the interval $(0,1)$. From \eqref{rownanieReliabImportPrec1} we obtain 
\begin{equation} \label{rownanieReliabImportPrec3}
    I_{h}(i;\vecp) = \bP\{\phi(1_i,\vecX) - \phi (0_i,\vecX)=1\}.
\end{equation}
From equations \eqref{rownanieReliabImportPrec1} and \eqref{rownanieReliabImportPrec3} we conclude that $I_h(i)$ can be interpreted as the probability that a system has a state, in which it is spoiled due to the $i$-th element being out of order.

\begin{example}[Birnbaum reliability importance - series]
Let us consider the series structure presented in Fig.~\ref{fig:seriesBirnbaum} composed of three independent components, where each component $c_1,c_2,c_3$ has a corresponding reliability $\vecp=(0.95,0.99,0.96)$.
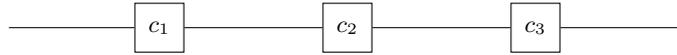
\begin{figure}[H]
    \begin{center}
        \tikzstyle{int}=[draw, minimum size=2em]
        \tikzstyle{init} = [pin edge={to-,thin,black}]
        \begin{tikzpicture}[node distance=2.5cm,auto,>=latex']
            \node [int] (A) {$c_1$};
            \node (B) [left of=A,node distance=2cm, coordinate] {A};
            \node [int] (C) [right of=A] {$c_2$};
            \node [int] (D) [right of=C] {$c_3$};
            \node [coordinate] (end) [right of=D, node distance=2cm]{};
            \path (B) edge node {} (A);
            \path (A) edge node {} (C);
            \path (C) edge node {} (D);
            \draw[-] (D) edge node {} (end) ;
        \end{tikzpicture}    
    \caption{Series structure} \label{fig:seriesBirnbaum}
    \end{center}
\end{figure}
Then, at any time $t$ the system reliability is equal to $h(\vecp) =\prod^{3}_{i=1}p_i=  0.90288$ and the Birnbaum reliability importance \eqref{eq:BirnbaumMeasure} for components $c_1,c_2,c_3$ is
\begin{equation*}
(B(1|\vecp),B(2|\vecp),B(3|\vecp))=(\prod_{\myatop{i=1}{i\neq 1}}^3p_i,\prod^{3}_{\myatop{i=1}{i\neq 2}}p_i,\prod^{3}_{\myatop{i=1}{i\neq 3}}p_i)=(0.9504,0.912,0.9405).
\end{equation*}
In the series system we may see that the component having the smallest reliability is the most meaningful for the system.
\end{example}

\begin{example}[Birnbaum reliability importance - parallel] \label{example:Birnbaumserie}
Let us consider the parallel structure presented in Fig.~\ref{fig:parallelBirnbaum} composed of three independent components
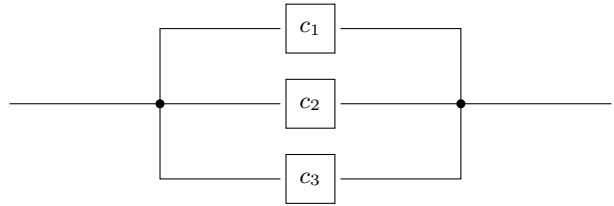
\begin{figure}[H]
    \begin{center}
        \tikzstyle{int}=[draw,fill=white, minimum size=2em]    
        \begin{tikzpicture}
            \node[int] (c1) at (0,0) {$c_1$};
            \node[int] (c2) at (0,-1) {$c_2$};
            \node[int] (c3) at (0,-2) {$c_3$};
            \draw[-] (-0.4,0)--(-2,0)--(-2,-1)--(-2,-2)--(-0.4,-2);
            \draw[-] (-0.4,-1)--(-4,-1);
            \draw[-] (0.4,0)--(2,0)--(2,-1)--(2,-2)--(0.4,-2);
            \draw[-] (0.4,-1)--(4,-1);
            
            \fill (-2,-1) circle (0.06cm) (2,-1) circle (0.06cm);
        \end{tikzpicture}    
    \caption{\label{fig:parallelBirnbaum}Parallel structure.} 
    \end{center}
\end{figure}\noindent
where components $c_1,c_2,c_3$ have the same reliabilities like in Example~\ref{example:Birnbaumserie}.
Then, the system reliability at time $t$ is equal to
\begin{equation*}
    h(\vecp) = \coprod^{3}_{i=1} p_i = 1 - \prod^{3}_{i = 1} (1-p_i) = 0.99998
\end{equation*}
and the Birnbaum reliability importance \eqref{eq:BirnbaumMeasure} for components  $c_1,c_2,c_3$ is
\begin{align*}
(B(1|\vecp),B(2|\vecp),B(3|\vecp))&=(\prod_{\myatop{i=1}{i\neq 1}}^3(1-p_i),\prod^{3}_{\myatop{i=1}{i\neq 2}}(1-p_i),\prod^{3}_{\myatop{i=1}{i\neq 3}}(1-p_i))\\
&=(0.0004,0.002,0.0005).
\end{align*}
In the parallel system we may see that the component having the greatest reliability is the most relevant for the system.
\end{example}

\subsection{Lifetime importance measure.} If $n$ components comprise the system, then we assume that for $t\geq 0$ and $i=1,2,...,n$ a stochastic process $X_i(\omega,t)$  defines the $i$-th component's state with $X_i(\omega,t)$ equal to $0$ or $1$, reliant on failure or functioning of the system at moment $t$, respectively. Let $\xi_i(\omega)=\inf\{t\in\Re^+:X_i(\omega,t)=0\}$--the life time of $i$th element and denote $Q_i(s)=\bP\{\omega;\xi_i(\omega)\geq s\}$. Assuming continuous life distribution of the $i$-th component $Q_i(t)=\bP\{\omega:X_i(\omega,t)=1\}$ and the structure $\phi$, at each moment $t$ there is defined the reliability of the structure by the adequate function $h(\vecQ(t))$ (see \eqref{reliabfun}). Based on these denotation we have the process of system states $X(\omega,t)=\phi(\vecX(\omega,t))$ and the system's reliability function $Q(t)$ can be derived.  Hence,
\begin{align}\label{RelFunSys}
    Q(t)=h(\vecQ(t))  &= h\big({Q}_1(t),{Q}_2(t),\ldots,{Q}_n(t)\big) \\ 
    & = \bP\{\omega:\phi(\vecX(\omega,t))=1\} = \bE\big[\phi(X(\omega,t)\big].
\end{align}
Let us calculate the density function $f(t)=-Q'(t)=-q(t)$ of the survival time distribution of the structure with the structure function $h$. 
\begin{align}
\label{survdenStrut}
f(t)=-\frac{d}{dt}Q(t)&\stackrel{\eqref{RelFunSys}}{=}-\left\langle \nabla h(\vecQ(t)),\vecq(t)\right\rangle \\
\nonumber &\stackrel{\eqref{Gradh}}{=}-\left\langle\vecI_h(\vecQ(t)),\vecq(t)\right\rangle=-\left\langle\bE[\vecdelta(\vecX)],\vecq(t)\right\rangle. 
\end{align} 
First, Birnbaum in 1968 introduced importance measures for fixed function of time $t$, while Barlow and Proschan in 1975 freed the measure from the time-dependence. They proposed probability that the system and $i$-th component failures coincide, which means that the $i$-th component impaired the system.

\begin{fact}\label{FactRelImp}
For $i=1,2,...,n$ let the $i$-th component have a distribution $F_i$, reliability $Q_i$ and density $f_i$. Then, the probability that system failure occurred at time t and was caused by a component $i$ is defined as

\begin{equation}\label{eq:lifetimeproof}
\frac{f_i(t) \cdot [h(1,\vecQ_{-i}(t)) - h(0,\vecQ_{-i}(t))] }{\left\langle \nabla h(\vecQ(t)),\vecf(t)\right\rangle}=\frac{f_i(t) \cdot I_h(i;\vecQ(t)) }{\sum_{k=1}^{n} f_k(t)\cdot I_h(k;\vecQ(t))}.
\end{equation}
\end{fact}

\begin{proof}
The probability of the system functioning at time $t$ if the $i$-th element is functioning and that the system is not functioning at time $t$ if the $i$-th element is not functioning is 
\begin{align}\nonumber P\big[\phi(1,\vecX_{-i}(t)) - \phi (0,\vecX_{-i}(t))=1\big] &= h(1,\vecQ_{-i}(t)) - h(0,\vecQ_{-i}(t))\\
\label{eqBarPro}		&=I_h(i;\vecQ(t)).
\end{align}
Therefore, the numerator in \eqref{eq:lifetimeproof} multiplied by $dt$ represents the probability that in the interval $[t, t+dt]$ the $i$-th component led to the failure of the system and the denominator multiplied by $dt$ stands for the probability that the system failed in the given interval \cite{BarPro1975:Importance}.
\end{proof} 
\begin{fact}
In consequence of equation \eqref{eqBarPro}, the probability of $i$ causing the system failure in time interval $[0,t]$, while the system failure occurs in the same period of time $[0,t]$ is
{\small
\begin{equation} \label{eqBarlow}
\frac{\int_0^t \cdot I_h(i;\vecQ(u))dF_i(u) }{\sum_{k=1}^{n} \int_0^t \cdot I_h(k;\vecQ(u))dF_k(t)}=    \frac{\int_{0}^{t}[h(1,\vecQ_{-i}(u)) - h(0,\vecQ_{-i}(u))] d F_i(u)}{\int_{0}^{t} \sum_{k=1}^{n}[h(1,\vecQ_{-k}(u)) - h(0,\vecQ_{-k}(u))]d F_k(u)}.
\end{equation}}
\end{fact}
When in equation \eqref{eqBarlow} $t \to \infty$, then it is a probability of $i$ leading the system to the total failure. In this regard, the denominator is equal to $1$. We assume that this limit is a definition of component importance.

\begin{definition}
As a consequence of \eqref{eqBarlow}, the probability of $i$ causing the system failure is denoted as
\begin{equation}
    I_h(i;\vecQ) =  \int\limits_{0}^{\infty} [h(1,\vecQ(t)) - h(0,\vecQ(t)(t))] d F_i(t)
\end{equation}
where $I_h(i;\vecQ)$ is precisely the lifetime importance measure of the $i$-th component. 
\end{definition}
\begin{fact}
Importance measure properties
\end{fact}
\begin{enumerate}
    \item $I_h(i;\vecQ) \in [0,1]$
    \item $I_h(i;\vecQ) \in (0,1)$ if $n \geq 2$
    \item $\sum_{i=1}^{n} I_h(i;\vecQ) = 1$
\end{enumerate}
However, Birnbaum in 1969 extended reliability importance of components, he was not able to free measure from time dependence though. Probability distribution $F_i(t) = P\{\xi_i \leq t\}$ was considered with assumption of each $i$-th component having a life length $\xi_i$ \cite{Bir1969:Multivariate}. Therefore, using this assumption and those from the beginning of this section, we have the lifetime importance measure given by
\begin{gather}
    I_h^i(t) = \bP\Big[\phi \big(1,\vecX_{-i}(t)\big) - \phi \big(0,\vecX_{-i}(t)\big) = 1 \Big] = \\
    = h(1,\vecQ_{-i}(t)) - h(0,\vecQ_{-i}(t)),
\end{gather}
what describes probability at time $t$ that the system is in the state $t$ in which the $i$-th component is crucial for the system.
If the $i$-th component is series or parallel to the system, then it has a corresponding formula to the structural importance case \cite{Xie1987:Importance}.

\subsection{Module importance} \label{sectionModuleImportance}

Multi-component and coherent system may be partitioned into modules, which in other words are sub-systems consisting of different components. As Birnbaum \cite{Bir1969:Multivariate} proposed, a module importance for fixed time with coherent structure $\phi$ is expressed by
\begin{equation}
    \phi (x) = \phi (x_1, x_2, ..., x_n) = x_1 \cdot \delta_{x_1} (\phi;x) +\mu_{x_1} \cdot (\phi;x)
\end{equation}
and coherent structure $\Psi (y)$ denoted as 
\begin{equation}
    \Psi (y) = \Psi(y_1,y_2,...,y_m),
\end{equation}
we may achieve the structure $\chi$, if in $\phi(x)$ an element $x_1$ is substituted by the coherent module $\Psi (y)$:
\begin{gather}
    \chi (y_1,...,y_m, x_2,..., x_n) = \phi [\Psi (y_1,...,y_m), x_2,...,x_n] = \phi [\Psi_1(y),x] = \\ \label{eqPsi}
    = \Psi(y) \cdot \delta_{x_1} \cdot[\phi;x] +\mu_{x_1} \cdot [\phi;x].
\end{gather}
From \eqref{eqPsi} we deduce that
\begin{equation}\label{eqPsi2}
   \begin{gathered} 
    \delta_{x_1}(\chi;y_1,...,y_m, x_2,..., x_n) = \\ = \chi(1,y_2,...,y_m, x_2,..., x_n) - \chi(0,y_2,...,y_m, x_2,..., x_n) =\\ = \delta_{x_1}(\phi;x) \cdot [\Psi(1,y) - \Psi(0_1,y)] = \delta_{y_1}(\Psi;y) \cdot  \delta_{x_1}(\phi;x).
\end{gathered} 
\end{equation}
If we base on equations \eqref{eqDerivDelta} and \eqref{eqPsi2}, then we obtain the importance of a module defined as
\begin{equation}
    I_{y_i} (\chi; y_1,...,y_m, x_2,..., x_n) = I_{x_i}(\phi;x) \cdot I_{y_i}(\Psi;y).
\end{equation}
For the system $\chi$ we may derive the importance of every component comprising the module $\Psi$ by repeating the procedure of substituting modules for components till none is left \cite{Bir1969:Multivariate}. 

Different definition of module importance of the coherent system was proposed by Barlow and Proschan in \cite{BarPro1975:Importance} (cf. \cite{Ber1985:NewImportance}).

\begin{definition}
For $n$ components let introduce a structure $\phi$ that is coherent, subset of {1,2,...,n} given by $M$ and its complement $M^C$, and coherent system $\chi$ comprised of components in $M$. Then, the module $(M,\chi)$ of the coherent system $\phi$ is defined as
\begin{equation}
    \phi (x) = \Psi [ \chi(x^M), x^{M^C} ],     
\end{equation}
where $x^{M^C}$ is a complement of a subset $M$.
\end{definition}
The module importance $I_h(M)$ is the probability of the module causing system failure.
\begin{theorem}
If $i \in M$ and $f$ denotes module's reliability function, then
\begin{equation}\label{eqBarModImp1}
    I_h(i) = \int\limits_{0}^{\infty} \big[ h(1^M, \bar{F}(t)) - h(0^M, \bar{F}(t)) \big] \cdot \big[ f(1_i,\bar{F}(t)) - f(0_i,\bar{F}(t)) \big] d \bar{F}_i(t)
\end{equation}
and
\begin{equation}\label{eqBarModImp2}
    I_h(M) = \sum_{i \in M} I_h(i)
\end{equation}
\end{theorem}

\begin{proof} \eqref{eqBarModImp1}
Probability of functioning of the system at time $t$, if and only if the module functions firmly, is represented by
\begin{equation*}
    h(1^M, \bar{F}(t)) - h(0^M, \bar{F}(t)) = P \big[ \phi(1^M, X(t)) - \phi(0^M, X(t)) = 1 \big],
\end{equation*}
while the probability of module functioning at time $t$, if and only if the component $i$ functions, is denoted as
\begin{equation*}
    f(1_i,\bar{F}(t)) - f(0_i,\bar{F}(t)) = P \big[ \chi(1_i, X(t)) - \chi(0_i, X(t)) = 1 \big],
\end{equation*}
In the system with modules, component $i$ may only cause system failure through module failure, hence
\begin{gather*}
    \sum_{i \in M} I_h(i) = \int\limits_{0}^{\infty} \big[ h(1^M, \bar{F}(t)) - h(0^M, \bar{F}(t)) \big] \cdot \sum_{i \in M} \big[ f(1_i,\bar{F}(t)) - f(0_i,\bar{F}(t)) \big] d F_i(t) = \\ = - \int\limits_{0}^{\infty} \big[ h(1^M, \bar{F}(t)) - h(0^M, \bar{F}(t)) \big] \frac{d}{dt} f(\bar{F}(t))dt = I_h(M)
\end{gather*}
\end{proof}
\begin{note}
Definition of the module importance proposed by Birnbaum is slightly different from the one introduced by Barlow and Proschan. In Birnbaum's definition importance of the module's component is equal to the component's importance for the module multiplied by the importance of the module for the system. This is not consistent with Barlow and Proschan definition due to the fact that for each $x$ expression $r(x)=s(x)\cdot u(x)$ doesn't imply 
\begin{equation*}
    \int_{a}^{b} r(x)dx = \int_{a}^{b} s(x)dx \cdot \int_{a}^{b} u(x)dx.
\end{equation*}
\end{note}

\begin{lemma} \label{LemmaModIm1}
If a component $i$ is serial to the system, then the importance $I_h(i)$ increases in $F_i(t)$ and $\bar{F}_j(t)$ when $i\neq j$. Otherwise, if component $i$ is parallel to the system, then the importance $I_h(i)$ decreases.
\end{lemma}
\begin{proof}
With assumption that component $i$ is serial to the system, we obtain
\begin{equation*}
    I_h(i) = \int_{0}^{\infty} h(1_i, \bar{F}(t)) d F_i(t),
\end{equation*}
while $h(0_i,\bar{F}(t)) = 0 $ due to the hypothesis. Since $h(1_i,p)$ increases in each $p$, $I_h(i)$ increases in $\bar{F}_j(t)$, if $i \neq j $. Moreover, $h(1_i,\bar{F}(t)$ decreases in $t$, therefore $I_h(i)$ increases in $F_i(t)$.
\end{proof}

\begin{lemma} \label{LemmaModIm2}
If component $i$ is serial or parallel to the system and all components have the same distribution $F$, then for $i \neq j$ we obtain $I_h(i) \geq I_h(j)$.
\end{lemma}
\begin{proof}
If we assume that $i$ is serial to the system and use the fact that components are stochastically alike, $I_h(k)$ may be treated as the permutations' proportion from $1$ to $n$ corresponding to the failure of the system by cause of $k$. Hence, computation of $I_h(k)$ proceeds with the interchange of $j$ and $i$ in each permutation. This calculation method shows that the number of permutations, in which the failure is caused by $i$, is not smaller than the number of permutations in which the failure is caused by $j$.
\end{proof}

By using lemmas \ref{LemmaModIm1} and \ref{LemmaModIm2} we may introduce the following theorem \ref{TheoremModIm}.
\begin{theorem} \label{TheoremModIm} 
If we assume that the $i$-th component is serial or parallel to the system and $t \geq 0$, $j \neq i$, then the true is $F_j(t) \leq F_i(t)$ and $I_h(j) \leq I_h(i)$.
\end{theorem}

\subsection{Structural importance} \label{sectionStructuralImportance}

At times we have to face the situation when information about component reliabilities are missing. In that case we have to consider the impact of various components to the system. And so, we define the structural importance. 

Measure introduced by Birnbaum \cite{Bir1969:Multivariate} requires specifying structure function equalities \eqref{structfun1}, \eqref{structfun2}, \eqref{structfun3}.

\begin{definition}
    \begin{enumerate}[a)]
    \item A component $c_i$ is indispensable for the $\phi$ structure at the vector of states $x$ when
    \begin{equation}
        \phi(1_j,x) - \phi(0_j,x) = \delta_j(x) = 1
    \end{equation}
    \item A component $c_i$ is indispensable at the vector of states $x$ for the functioning of the structure $\phi$ when
    \begin{equation}
       (1 - x_j) \cdot \delta_j(x) = 1
    \end{equation}
    \item A component $c_i$ is indispensable at the vector of states $x$ for the failure of the structure $\phi$ when
    \begin{equation}
       x_j \cdot \delta_j(x) = 1
    \end{equation}
    \end{enumerate}
\end{definition}
To clarify, if $c_i$ is indispensable at the state vector $x$, then it is equally indispensable for both functioning or failure, when coordinates of the state vector $x$ equal to $0$ or $1$.

Hence, the structural importance of a component $c_i$ for the functioning of $\phi$ is defined as
\begin{equation}\label{eqStrImp1}
    I_{\phi}(j,1) = 2^{-n}\sum_{(x)} (1-x_j) \cdot \delta_j(x),
\end{equation}
where the sum covers all $2^n$ unit cube's vertices. The structural importance of a component $c_i$ for failure of structure $\phi$ is defined as
\begin{equation}\label{eqStrImp2}
    I_{\phi}(j,0) = 2^{-n}\sum_{(x)} x_j \cdot \delta_j(x)
\end{equation}
and the structural importance of a component $c_i$ for the structure $\phi$ is defined as
\begin{equation}\label{eqStrImp3}
    I_{\phi}(j) = I_{\phi}(j,1) + I_{\phi}(j,0) = \sum_{(x)} \delta_j(x).
\end{equation}
To conclude, if a component $c_i$ is indispensable at the state vector $\vecx$ for functioning of structure $\phi$, then the component $c_i$ is indispensable at $(1,\vecx_{-j})$ for failure, meanwhile, if a component $c_i$ is indispensable at the state vector $x$ for failure of structure $\phi$, then the component $c_i$ is indispensable at $(0,\vecx_{-j})$ for functioning. Due to communication between vertices at which $c_i$ is responsible for failure or functioning, the number of each type of vertices is the same. Hence, from equalities \eqref{eqStrImp1}, \eqref{eqStrImp2} and \eqref{eqStrImp3} follows
\begin{equation} \label{eqStrImp4}
    I_{\phi}(j,1) = I_{\phi}(j,0) = \frac{1}{2} I_{\phi}(j).
\end{equation}
From \eqref{eqStrImp4} we deduce that there is no purpose dividing the structural importance into the one for failure and the one for functioning, unlike the reliability importance.

When we consider continuous life distribution of components, we shall use the structural importance measure introduced by Barlow and Proschan. The importance of component $c_i$ proposed in fact \ref{FactRelImp} with assumption that all components have the same life distribution $F_1=F_2=...=F_n$, then in structure $\phi$, such an importance becomes the structural importance of component $c_i$ denoted as $I_{\phi}(i)$. By substituting $p$ for $\bar{F}_i(t)$ for $i=1,...,n$, we obtain
\begin{equation}\label{eqStrImp5}
    I_{\phi}(i) =  \int\limits_{0}^{1} [h(1_i,p) - h(0_i,p)] d p,
\end{equation}
where vector $(1,\vecp_{-i})$ has $1$ in the $i$-th position and $p$ everywhere else i.e. $\vecp_{-i}=\vec{p}_{-i}$.

To compute the structural importance presented by \citeauthor{BarPro1975:Importance}~(\citeyear{BarPro1975:Importance}), first we need to introduce some definitions.
\begin{definition}
    \begin{enumerate}[a)]
        \item A set of elements that allow proper operating of the system is called a path set. If the path set is irreducible, then it is called a minimal path set.
        \item At the same time, a set of elements that can by their own effect failure of the system is called a cut set. If the cut set is irreducible, then it is called a minimal cut set.
        \item A vector $(1_i,x)$, which fulfills conditions of $\phi(0_i,x)=0$ and $\phi(1_i,x)=1$, is called a critical path vector for the $i$-th component. Hence, for the $i$-th component a critical path set is denoted as
        \begin{equation*}
            \{i\} \cup \{j | x_j = 1, i \neq j\}.
        \end{equation*}
    \end{enumerate}
\end{definition}
It means that functioning of the system or its failure is determined by a component $c_i$. A critical path vector for a component $c_i$ with size $r$ can be presented as
\begin{equation*}
    1 + \sum_{i \neq j} x_j= r, \qquad r = 1,2,...,n.
\end{equation*}
Therefrom, we may introduce a number of critical path vector $n_r(i)$ for $i$-th component of size $r$ specified as
\begin{equation*}
    n_r(i) =  \sum_{\sum_{i\neq j}x_j=r-1} [ \phi(1,\vecx_{-j}) - \phi(0,\vecx_{-j}) ]
\end{equation*}
Hence, the structural importance $I_{\phi}(i)$ may be expressed as the number of critical path vectors $n_r(i)$.
\begin{theorem}
\begin{equation}\label{eqStrImp6}
    I_{\phi}(i) = \sum_{r=1}^{n} n_r(i) \cdot \frac{ (n-r)!(r-1)!}{n!}
\end{equation}
\end{theorem}
\begin{proof}
If we merge equations \eqref{eqStrImp5} and \eqref{eqStrImp6}, we obtain
\begin{gather}
    I_{\phi}(i) = \int\limits_{0}^{1} [h(1_i,p) - h(0_i,p)] d p = \\
    = \int\limits_{0}^{1} \Big[ \sum_{x}[\phi(1_i,x) - \phi(0_i,x)] \cdot p^{\sum_{j\neq i} x_j} \cdot (1-p)^{ n-1 - \sum_{j\neq i} x_j} \Big]dp = \\
    = \int\limits_{0}^{1} \sum_{r=1}^{n} n_r(i) \cdot p^{r-1} (1-p)^{n-r}dp =  \sum_{r=1}^{n} n_r(i) \cdot \frac{ (n-r)!(r-1)!}{n!}.\label{eqStrImp7}
\end{gather}
\end{proof}
Equation \eqref{eqStrImp6} can be expressed as
\begin{equation}\label{eqStrImp8}
     I_{\phi}(i) = \frac{1}{n}  \sum_{r=1}^{n} n_r(i) \tbinom{n-1}{r-1}^{-1},
\end{equation}
where the numerator $n_r(i)$ stands for the critical path vectors with size $r$ and the denominator stands for the number of results with precisely $r-1$ components functioning among the exactly $n-1$ components, without component $i$. It means that the $i$-th component's structural importance is in other words the average probability that for the $i$-th component the vector is the critical path vector.

Expression \eqref{eqStrImp7} can be also translated into
\begin{equation}\label{eqStrImp9}
    I_{\phi}(i) = \int\limits_{0}^{1} \Big[ \sum_{r=1}^{n}  n_r(i) \cdot \tbinom{n-1}{r-1}^{-1} \tbinom{n-1}{r-1} \cdot (1-p)^{n-r} \cdot p^{r-1}\Big]dp
\end{equation}
where $\tbinom{n-1}{r-1} \cdot (1-p)^{n-r} \cdot p^{r-1} $ is the probability that $r-1$ among $n-1$ elements, omitting the $i$-th one, function. Furthermore, $n_r(i) \cdot \tbinom{n-1}{r-1}^{-1}$ is the probability that functioning components $i$ and $r-1$ comprise the critical path set for the component $i$. Hence, equation \eqref{eqStrImp9} stands for the probability of $i$ causing the system failure. Integrating it over $p$ means that the component reliability $p$ has a uniform distribution $p \sim \mathcal{U}(0,1)$.

If we compare Barlow and Proschan structural importance
\begin{equation}\label{eqStrImp10}
    I_{\phi}(i;\vecp) =  \int\limits_{0}^{1} [h(1,\vec{p}_{-i}) - h(0,\vec{p}_{-i})] d p,
\end{equation}
with Birnbaum structural importance
\begin{equation}\label{eqStrImp11}
    B(i;\overrightarrow{.5}) = I_{\phi}(i;\overrightarrow{.5}) = \frac{\partial h(p)}{\partial p_i}\Bigg\rvert _{p_1=...=p_n=\frac{1}{2}} = h\big(1,\overrightarrow{.5}_{-i}\big) - h\big(0,\overrightarrow{.5}_{-i}\big)
\end{equation}
we see that Birnbaum sets $p=\tfrac{1}{2}$ in order to compute the difference $h(1,\vecp_{-i}) - h(0,\vecp_{-i})$, while Barlow and Proschan compute this difference for $p \in [0,1]$.

Moreover, from equation \eqref{eqStrImp11} we can deduce that
\begin{equation*}
    B(i;\vecp) = I_{\phi}(i;\vecp) = \sum_{x} \frac{1}{2^{n-1}} \cdot [\phi(1,\overrightarrow{.5}_{-i}) - \phi(0,\overrightarrow{.5}_{-i})].
\end{equation*}
Hence, Birnbaum structural importance can be given as
\begin{equation}\label{eqStrImp12}
     I_{\phi}(i;\overrightarrow{.5}) = \sum_{r=1}^{n} \frac{n_r(i)}{2^{n-1}}.
\end{equation}
If we compare expressions \eqref{eqStrImp6} and \eqref{eqStrImp12}, we can see that in $I_{\phi}(i)$ the number of critical vector path $n_r(i)$ has a weight $ (n-r)!(r-1)!/n!$, meanwhile Birnbaum uses the same weight $1/(2^{n-1})$ everywhere. Due to behavior of $ (n-r)!(r-1)!/n!$ for different $n$ we deduce that only very large or very small critical paths may reach the greatest weight \cite{BarPro1975:Importance}.

\begin{example}[{\small Structural importance - series / parallel structure}]
Let consider a structure of $n$ elements, which $k$ are in series and $n-k$ are parallel. This example concerns a structure of five components $c_1,c_2,c_3,c_4,c_5$, $n=5$ and $k=2$ presented in figure \ref{fig:BirStruct}.
\begin{figure}[H]
    \begin{center}
        \tikzstyle{int}=[draw,fill=white, minimum size=2em]    
        \begin{tikzpicture}
            \node[int] (c1) at (-4.5,-1) {$c_1$};
            \node[int] (c2) at (-3,-1) {$c_2$};
            \node[int] (c3) at (0,0) {$c_3$};
            \node[int] (c4) at (0,-1) {$c_4$};
            \node[int] (c5) at (0,-2) {$c_5$};

            \draw[-] (-0.4,0)--(-1.5,0)--(-1.5,-1)--(-1.5,-2)--(-0.4,-2);
            \draw[-] (0.4,0)--(1.5,0)--(1.5,-1)--(1.5,-2)--(0.4,-2);
            \draw[-] (0.4,-1)--(3,-1);
            \draw[-] (-0.4,-1)--(-2.6,-1);
            \draw[-] (-3.4,-1)--(-4.1,-1);
            \draw[-] (-4.9,-1)--(-6.5,-1);
            
            \fill (-1.5,-1) circle (0.06cm) (1.5,-1) circle (0.06cm);
        \end{tikzpicture}    
    \caption{Series and parallel structure} \label{fig:BirStruct}
    \end{center}
\end{figure}
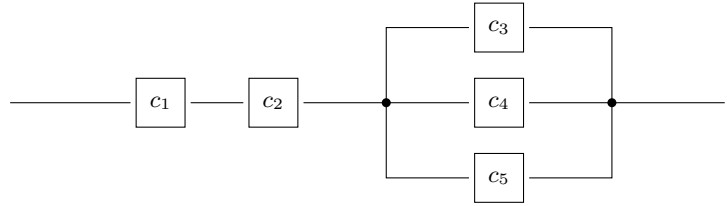
Reliability of each component $c_i$ is unknown, however we may derive the system reliability for different $p_i$
\begin{equation*}
    h(\vecp) = p_1 \cdot p_2 \cdot \big[ 1 - (1-p_3) \cdot (1-p_4) \cdot (1-p_5)  \big].
\end{equation*}
The structural importance's assumption is that the reliabilities $p_1 = p_2 = ... = p_n=p$ are identical. Since 
\begin{equation*}
    I_B(j;\vecp) = \bD_{p_j}h(\vecp),
\end{equation*}
we may derive formulas for the structural importance for each component
\begin{align} 
 \label{eq:delta1}&I_B(j)  = \prod^k_{\substack{i=1 \\ i\neq j}} p_i \big[ 1 - \prod^n_{m=k+1} (1-p_m) \big] \text{\quad for $j=1,...,k$,}  \\
  \label{eq:delta2}&I_B(j) = \prod^k_{i=1} p_i \prod^n_{\substack{m=k+1 \\ m\neq j}} (1-p_m)  \text{\quad for $j=k+1,...,n$.}\\
\intertext{For Birnbaum case, each reliability $p_i =\frac{1}{2}$, hence from \eqref{eq:delta1} and \eqref{eq:delta2} we obtain} 
\nonumber & I_B(j;\overrightarrow{.5}) = 2^{-(k-1)} - 2^{-(n-1)}         \text{\quad for $ i=1,\ldots,k$,} \\
\nonumber & I_B(j;\overrightarrow{.5}) = 2^{-(n-1)} 									 \text{\quad for $i=k+1,\ldots,n$.} \\
\intertext{Therefore, for structure in figure \ref{fig:BirStruct} with $k=2$ and $n=5$, we have}
\nonumber & I_B(1;\overrightarrow{.5}) = I_B(2;\overrightarrow{.5}) = 2^{-1} - 2^{-4} = 0.4375 \\
\nonumber & I_B(3;\overrightarrow{.5}) = I_B(4;\overrightarrow{.5}) = I_B(5;\overrightarrow{.5}) = 2^{-4} = 0.0625
\end{align}
We can see that the components in series have much greater importance than the components in parallel.
\end{example}

\begin{example}[Minimal path and cut sets]
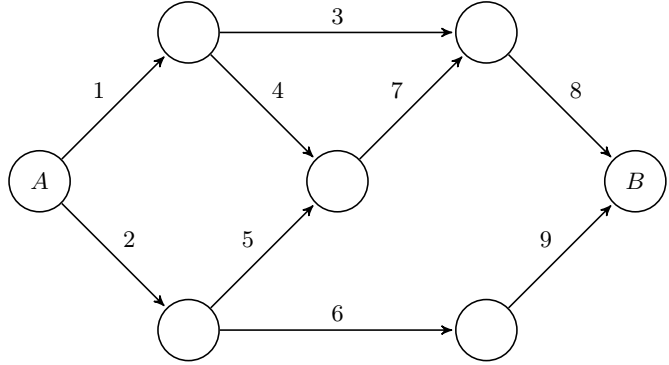
\begin{figure}[H]
    \begin{center}
        \begin{tikzpicture}[->,>=stealth',shorten >=1pt,auto,node distance=2.8cm,semithick]
            \node[state] (A)                    {$A$};
            \node[state] (C) [above right of=A] {};
            \node[state] (F) [below right of=A] {};
            \node[state] (E) [below right of=C] {};
            \node[state] (D) [above right of=E] {};
            \node[state] (G) [below right of=E] {};
            \node[state] (B) [below right of=D] {$B$};
        
            \path   (A) edge node {1} (C)
                        edge node {2} (F)
                    (C) edge node {3} (D)
                        edge node {4} (E)
                    (D) edge node {8} (B)
                    (E) edge node {7} (D)
                    (F) edge node {5} (E)
                        edge node {6} (G)
                    (G) edge node {9} (B);
        \end{tikzpicture}
        \caption{Graph $G_{A,B}$}
        \label{graph}
    \end{center}
\end{figure}
Let's consider the structure of order $10$ with edges of graph $G_{A,B}$ as elements of the system. Every set of edges connecting vertices $A$ and $B$ is a path and every set of edges, when removed, disconnecting vertices $A$ and $B$ is a cut \cite{kordecki}. Directed graph $G_{A,B}$ represents simple example of the network connecting two nodes (A and B) that is often used in examining reliability of computer networks. In order to define structure function $\phi (x)$, determining the minimal path and cut sets is obligatory.

\vspace{-1ex}\small
\hspace{-2em}\begin{minipage}{.403\linewidth}
\vspace{-3ex}
\begin{table}[H]
        \centering
        {\small\caption{\label{tab:minpathset}Minimal path set}
        \begin{tabular}{|c|c|} 
            \hline
            Path & Elements    \\ [0.5ex] 
            \hline
            1 & 1 3 8          \\ 
            2 & 1 4 7 8        \\
            3 & 2 5 7 8        \\
            5 & 2 6 9          \\
            \hline
        \end{tabular}}    
\end{table}
\end{minipage}
\hspace{.5em}
\begin{minipage}{.6\linewidth}
The minimal path and cut sets presented in tables \ref{tab:minpathset} and \ref{tab:mincutset} respectively, allow to determine the structure function. Graph $G_{A,B}$ is described by four minimal path series structures
\begin{align*}
    \rho_1(\vecx) &=\prod_{i\in\{1,3,8\}}\hspace{-1em}x_i
		          &    \rho_2(\vecx) &=\prod_{i\in\{1,4,7,8\}}\hspace{-1em}x_i\\
    \rho_3(\vecx) &=\prod_{i\in\{2,5,7,8\}}\hspace{-1em}x_i
		          &     \rho_4(\vecx) &=\prod_{i\in\{2,6,9\}}\hspace{-1em}x_i
\end{align*}
\end{minipage}

\hspace{-2em}\begin{minipage}{.403\linewidth}
\begin{table}[H]	
   \centering
         {\small\caption{\label{tab:mincutset}Minimal cut set} 
        \begin{tabular}{|c|c|} \hline 
            Cut & Elements \\ [0.5ex] 
            \hline
            1 & 1 2        \\ 
            2 & 1 5 6      \\
            3 & 1 5 9      \\
            4 & 1 6 7      \\
            5 & 1 6 8      \\
            6 & 1 7 9      \\
            7 & 2 3 4      \\
            8 & 2 3 7      \\
            9 & 2 8        \\
            10 & 3 4 5 6    \\
            11 & 3 4 5 9   \\
            12 & 3 6 7     \\
            13 & 3 7 9     \\
            14 & 6 8       \\
            15 & 8 9       \\
            \hline
        \end{tabular}}     
\end{table}
\end{minipage}
\hspace{.5em}
\begin{minipage}{.6\linewidth}
and by fifteen minimal cut parallel structures
\begin{align*}
    \kappa_1(\vecx) &= x_1 \amalg x_2     &\kappa_2(\vecx) &=\coprod_{i\in\{1,5,6\}}x_i\\
    \kappa_3(\vecx) &=\coprod_{i\in\{1,5,9\}}x_i
		            &    \kappa_4(\vecx) &=\coprod_{i\in\{1,6,7\}}x_i\\
    \kappa_5(\vecx) &=\coprod_{i\in\{1,6,8\}}x_i 
		            &    \kappa_6(\vecx) &=\coprod_{i\in\{1,7,9\}}x_i\\
    \kappa_7(\vecx) &=\coprod_{i\in\{2,3,4\}}x_i 
								&    \kappa_8(\vecx) &=\coprod_{i\in\{2,3,7\}}x_i\\
    \kappa_9(\vecx) &= x_2 \amalg x_8 %
		            &    \kappa_{10}(\vecx) &=\coprod_{i\in\{3,4,5,6\}}x_i\\
    \kappa_{11}(\vecx) &=\coprod_{i\in\{3,4,5,9\}}x_i
									 &     \kappa_{12}(\vecx) &=\coprod_{i\in\{3,6,7\}}x_i\\
    \kappa_{13}(\vecx) &= x_3 \amalg x_7 \amalg x_9 &    \kappa_{14}(\vecx) &= x_6 \amalg x_8 \\
    \kappa_{15}(\vecx) &= x_8 \amalg x_9 &&
\end{align*}\normalsize
\end{minipage}
The structure functions, if at least one of the minimal paths functions, can be presented as a parallel structure of minimal path series structure\vspace{-1ex}
\begin{equation*}
    \phi(\vecx) = \coprod_{i=1}^{r} \rho_i(\vecx) = 1 - \prod_{i=1}^{r} \big[ 1-\rho_i(\vecx) \big],
\end{equation*}

where $r$ is a number of minimal paths of graph $G_{A,B}$. Hence, the structure function of the graph $G_{A,B}$ can be written in the form of  
\begin{align*}
    \phi(\vecx) &= \coprod_{i=1}^4\rho_i(\vecx)= 1 - \prod_{j=1}^4(1-\rho_j(\vecx))\\
    & = 1 - (1 - \prod_{i\in\{1,3,8\}}\hspace{-1em}x_i)(1 - \prod_{i\in\{1,4,7,8\}}\hspace{-1em}x_i)(1 - \prod_{i\in\{2,5,7,8\}}\hspace{-1em}x_i)(1 -\prod_{i\in\{2,6,9\}}\hspace{-1em}x_i)
\end{align*}
and the structure function $\phi(\vecx)$ is exactly equal to 
\begin{equation*}
\begin{split}
\phi(\vecx) &= x_1^2 x_2 x_3 x_4 x_5 x_7^2 x_8^3 - x_1^2 x_2^2 x_3 x_4 x_5 x_6 x_7^2 x_8^3 x_9 - x_1 x_2 x_4 x_5 x_7^2 x_8^2\\
 &\quad - x_1^2 x_3 x_4 x_7 x_8^2 - x_1 x_2 x_3 x_5 x_7 x_8^2 + x_1 x_2^2 x_4 x_5\cdot x_6 x_7^2 x_8^2 x_9 \\
 &\quad + x_1^2 x_2 x_3 x_4 x_6 x_7 x_8^2 x_9 + x_1 x_2^2 x_3 x_5 x_6 x_7 x_8^2 x_9  + x_1 x_3 x_8 + x_1 x_4 x_7 x_8\\
 &\quad  +  x_2 x_5\cdot x_7 x_8 - x_1 x_2 x_3 x_6 x_8 x_9 - x_1 x_2 x_4 x_6 x_7 x_8 x_9 + x_2 x_6 x_9 - x_2^2 x_5 x_6 x_7 x_8 x_9.
\end{split}
\end{equation*}
Moreover, the structure function can be also presented in the form of the series structure of the minimal cut parallel structures
\begin{equation*}
    \phi(\vecx) = \prod_{i=1}^{c} \kappa_i(\vecx),
\end{equation*}
where $c$ is a number of minimal cuts of the graph $G_{A,B}$. If at least one of the minimal cuts fails, the structure fails as well. $G_{A,B}$ graph's structure function can be written in the short form as
\begin{equation*}
    \phi(\vecx) =\prod_{i=1}^{14} \kappa_i(\vecx).
\end{equation*}
\end{example}

The structural importance measure presented by Birnbaum in 1968 and then by Barlow and Proschan in 1975 was once independently developed in the field of game theory (v. Appendix~\ref{PowerIndex}) by Shapley and Shubik in 1954(v.  \citeauthor{Sha53:value}~(\citeyear{Sha53:value})) and \citeauthor{Ban1965:voting}~(\citeyear{Ban1965:voting}), respectively (v. \citeauthor{Ram1990:Simple}~(\citeyear{Ram1990:Simple}). 

\subsection{\label{GameImpMeasure}Importance measure based on multilateral stopping problem.}
The basis for the description of binary systems is the structure function described in appendix xxx. We consider semi-coherent structures, which means that the structure function has properties identical to the function aggregating players' decisions in multi-person decision problems considered in the work of \citeauthor{szayas95:voting}~(\citeyear{szayas95:voting}). Multi-player decision problems assume that each game participant has a preference function based on a scalar function defined on the states of a certain process. If the elements of the structure are assigned to conservators (hypothetical players) who take care of the condition of these elements so that they fulfill their functions properly, the mentioned function can estimate profits and losses resulting from the state of the element. In principle, this condition should be good, allowing the function of the element, or bad - excluding the element from functioning. However, in reality, it is the diagnostician who decides when to perform maintenance or replacement (and bear the cost of it), and only sometimes a failure introduces a forced repair. An element in a system usually lowers its efficiency (e.g., mating components in a driveline may need lubrication to reduce friction, which results in increased energy expenditure and lower system efficiency), but the maintenance downtime is wasted and cannot always be managed. The operating conditions of the system make it possible to determine the correct payment function (cost) for each maintenance technician. Each of the n conservators, observing the states on which its payment depends, decides whether to order a maintenance break or to carry out uninterrupted operation. For safety reasons and the structure of the system, it is clear whether such a decision of a single observer is effective - it can start work when the system is stopped, and the stoppage requires the consensus of conservators from some critical path.

To analyze the effects of action, we will use the model of the following antagonistic game with elements of cooperation, which are defined by the function of the structure.

Following the results of the author and Yasuda~\cite{szayas95:voting} the multilateral stopping of a Markov chain problem can be described in the terms of the notation used in the non-cooperative game theory (see \cite{nas51:noncoop}, \cite{dresh81:games}, \cite{Mou1982:GT}, \cite{Owe2013:GT}). To this end the process and utilities of its states should be specified.
\begin{definition}[{\small ISS-Individual Stopping Strategies}]
Let $(\vecX_n,\cF_n,{\bP}_x)$, $n=0,1,2,\ldots ,N$, be a homogeneous Markov chain with the state space $(\bbE,\cB)$.
\begin{itemize}
 \item The players are able to observe the Markov chain sequentially. The horizon can be finite or infinite: $N\in\bbN\cup\{\infty\}$. 
\item Each player has their utility function $f_i: \bbE\rightarrow \Re $, $i=1,2,\ldots ,p$, such that ${\bE}_x|f_i(\vecX_1)|<\infty $ and the cost function $c_i: \bbE\rightarrow \Re $, $i=1,2,\ldots ,p$. 
\item If the process is not stopped at moment $n$, then each player, based on $\cF_n,$ can declare independently their willingness to stop the observation of the process.
\end{itemize}
\end{definition}
\begin{definition}[see~\cite{YasNakKur1982:Multivariate}] 
An individual stopping strategy of the player $i$ (ISS) is the sequence of random variables $\{\sigma _n^i\}_{n=1}^N$, where $\sigma_n^i:\Omega \rightarrow \{0,1\}$, such that $\sigma _n^i$ is $\cF_n$-measurable.
\end{definition}

The interpretation of the strategy is following. If $\sigma _n^i=1$, then player $i$ declares that they would like to stop the process and  accept the realization of $X_n$.

\begin{definition}[SS--Stopping Strategy (the aggregate function).] Denote 
\[
\sigma ^i=(\sigma _1^i,\sigma _2^i,\ldots ,\sigma _N^i)
\]
 and let $\scrS^i$ be the set of ISSs of player $i$, $i=1,2,\ldots ,p$. Define $\scrS=\scrS^1\times \scrS^2\times \ldots\times \scrS^p$. The element $\sigma =(\sigma ^1,\sigma ^2,\ldots ,\sigma ^p)^T\in \scrS$ will be called the stopping strategy (SS). 
\end{definition}
The stopping strategy $\sigma \in \scrS$ is a random matrix. The rows of the matrix are the ISSs. The columns are the decisions of the players at successive moments. The factual stopping of the observation process, and the players realization of the payoffs is defined by the stopping strategy exploiting $p$-variate logical function. 

Let $\delta :\{0,1\}^p\rightarrow \{0,1\}$ be the aggregation function. In this stopping game model the stopping strategy is the list of declarations of the individual players. The aggregate function $\delta$ converts the declarations to an effective stopping time.

\begin{definition}[An aggregated SS]
A stopping time $\tau_\delta (\sigma )$ generated by the SS $\sigma \in \scrS$ and the aggregate function $\delta $ is defined by 
\[
\tau_\delta (\sigma )=\inf \{1\leq n\leq N:\delta (\sigma _n^1,\sigma _n^2,\ldots ,\sigma _n^p)=1\}
\]
$(\inf (\emptyset )=\infty )$. Since $\delta $ is fixed during the analysis we skip index $\delta $ and write $\tau(\sigma )=\tau_\delta (\sigma )$. 
\end{definition}
\begin{definition}[Process and utilities of its states]
\begin{itemize}
\item $\{\omega \in \Omega : \tau_\delta (\sigma )=n\} =\bigcap\nolimits_{k=1}^{n-1}\{\omega \in \Omega : \delta (\sigma _k^1,\sigma_k^2,\ldots,\sigma _k^p)=0\} \cap \{\omega \in \Omega :\delta (\sigma_n^1,\sigma _n^2,\ldots,\sigma _n^p)=1\}\in \cF_n$;
\item $\tau_\delta (\sigma )$ is a stopping time with respect to $\{\cF_n\}_{n=1}^N$. 
\item For any stopping time $\tau_\delta (\sigma )$ and $\fri\in \{1,2,\ldots ,p\}$ the payoff of player $\fri$ is defined as follows (cf. ~\cite{shi78:book}):
\[
f_i(X_{\tau_\delta (\sigma )})=f_i(X_n)\one_{\{\tau_\delta (\sigma )=n\}}+\limsup_{n\rightarrow \infty }f_i(X_n)\one_{\{\tau_\delta (\sigma )=\infty\}}.
\]
\end{itemize}
\end{definition}
\begin{definition}\label{equdef}[An equilibrium strategy ({\rm cf.~\cite{szayas95:voting}})] Let the aggregate rule $\delta $ be fixed. The strategy
${}^{*}\!\sigma =({}^{*}\!\sigma^1,{}^{*}\!\sigma ^2,\ldots ,{}^{*}\!\sigma ^p)^T\in \scrS$ is an equilibrium strategy with respect to $\delta $ if for each $\fri\in \{1,2,\ldots ,p\}$ and any $\sigma^i\in \scrS^i$ we have
\begin{equation}\nonumber 
v_i(\vecx)={\bE}_x[ f_i(\vecX_{\tau_\delta ({}^{*}\!\sigma)})+\sum_{k=1}^{\tau_\delta ({}^{*}\!\sigma)}c_i(\vecX_{k-1})]
\leq  {\bE}_x[f_i(\vecX_{\tau_\delta({}^{*}\!\sigma(i))})+\sum_{k=1}^{\tau_\delta({}^{*}\!\sigma(i))}c_i(\vecX_{k-1})]. 
\label{defequ}
\end{equation}
\end{definition}
\begin{definition}\label{VGIequdef}[Voting Game Importance] Let the aggregate rule $\delta=h $ be fixed and the strategy
${}^{*}\!\sigma =({}^{*}\!\sigma^1,{}^{*}\!\sigma ^2,\ldots ,{}^{*}\!\sigma ^p)^T\in \scrS$ be an equilibrium strategy with respect to $\delta $. The voting game importance of the elements is the component of 
\begin{equation}\nonumber
\textbf{VGI}=\frac{\bE_{\vecQ^0}\vecv(\vecX)}{\bE<\vecv(\vecX),\vecQ^0>}. 
\end{equation}
\end{definition}

The measure of significance of a structure element introduced in this way takes into account its role in the structure by the aggregation function $ h $, it is normalized in the sense that the measures of all elements sum up to $ 1 $. It takes into account the external loads of elements, the cost of maintenance and repairs. Its use requires in-depth knowledge of the system and its components, which is a significant obstacle in its introduction into diagnostic practice. The hardest part is figuring out the payout functions (cost, risk, profit). The simplified version of the method may include in the payout functions only the operating risk with components in a condition requiring maintenance or repair, which is usually associated with less safety.

\section{Concluding remarks}
\subsection{Summary}
Ensuring the reliability and secure performance of the simple as well as complex systems has an indisputable significance in system analysis. Wherefore, the aim of the research was to answer the question how to recognize the most influential elements of the system so as to improve its reliability. This paper has demonstrated several approaches to the concept of importance measure depending on parameters and assumptions characterizing the system. The new approach is proposed in section~\ref{GameImpMeasure}.

In this paper we have considered binary systems. Their extension in the form of multistate systems is subject of another paper. In addition, the assumption was their coherence. Limitations and assumptions of coherent system for binary systems have been presented. For two-state systems various importance measures have been introduced and also the concept of the module importance, that can be applied to any more complex system. We have looked into case when only structure of the system was known (structural importance measure), case when the system was dependent on both reliability of components and structure of the system (reliability importance measure), and case, when the measure was dependent on the lifetime distribution of the components and the system structure (lifetime importance measure). The measures of importance have been based on Barlow and Proschan and Birnbaum's studies. The problem of choosing the proper importance measure was shown, e.g. due to the inconsistent behavior for different structures of the system. In addition, the relationship between importance measures in the reliability theory and power indices in the game theory have been discussed in the paper.

This analysis showed that the importance measures first introduced by Birnbaum in 1968 became the foundation for further search of more convenient and versatile definitions of the importance of components in the system reliability. Since then, research has expanded in different directions but till nowadays importance evaluation of highly complex structures such as networks may cause many computational problems. Besides, restrictions regarding coherence may exclude examination of certain systems. Therefore, this subject is under constant exploration.

\subsection{Exploratory importance measure research.}  There are many quantities estimated in probabilistic risk assessments (PRAs) to index the level of plant safety. If the PRA is to be used as a risk management tool to assist in the safe operation of the plant, it is essential that those elements of the plant design and its mode of operation that have the greatest impact on plant safety be identified. These elements may be identified by performing importance calculations. There are certain decisions that must be made before the importance calculation is carried out. The first is the definition of the events for which importance is to be evaluated; that is, to what level of resolution the analysis is to be performed. The second decision that must be made--and the major subject of this paper--is the choice of importance measure. Many measures of importance have been proposed; this discussion is restricted to three: the risk achievement (or degradation) worth, the risk reduction worth, and criticality importance. In \citeauthor{EPRINP3912SRVol2}~(\citeyear{EPRINP3912SRVol2}) these measures of importance are defined, their interrelationships are discussed, and a generalized importance measure is introduced. The use of these three measures is compared and their advantages and disadvantages are discussed. 

\subsection{Important direction of further investigations.} When interpreting component importance (v. \citeauthor{WuCoo2013:Cost-based}(\citeyear{WuCoo2013:Cost-based})), concluded that the importance of a component should depend on the following factors:
\begin{enumerate}
\item The location of the component in the system.
\item The reliability of the component.
\item The uncertainty in the estimate of the component reliability and related cost.\label{StatisticalImport}
\item The costs of maintaining this component in a given time interval $(0, t)$.
\end{enumerate}
(v. also \citeauthor{RauBarHoy2021:book}~(\citeyear{RauBarHoy2021:book})). The factor \eqref{StatisticalImport} highly depends on the statistical method implemented in the analyzes of exploratory data analyzes. Due to source of the data, the role of structure of the system to the reliability of it, the importance measure should take these elements into accounts. We are not observing the hidden state of the system directly and the information taken from the sensors should by interpreted and evaluated to infer on the hidden state of the elements and the system. The details of the construction needed, based on the results by \citeauthor{Sza2020:Rationalization}~(\citeyear{Sza2020:Rationalization}), are subject of a paper under editorial process.

\medskip


\vspace{6pt} 
\authorcontributions{Both authors equally contributed to the  conceptualization, methodology, formal analysis, investigation and writing--original draft preparation. Małgorzata Średnicka is responsible for the description of the importance measure concepts, examples, visualisation (v. \cite{Sre2019:MSc}) and Krzysztof J. Szajowski is responsible for the project conceptualization and its administration. 
}
\funding{This research received no external funding. 
}


\conflictsofinterest{The authors declare no conflict of interest.
} 
\medskip
\leftline{\large\bf Appendices}
\appendix
\section{\label{StruFun}Structure functions. } To study the relationship between the reliability of the components of a structure and the reliability of the structure itself, one has to know how the performance or failure of various components affect the performance or failure of the structure. We do this with the help of Boolean functions. In the reliability literature, Boolean functions are called structure functions.  Structure functions serve as a conceptual model on which the theory of reliability is largely based. The state of the system is assumed to depend only on the states of the components. We shall distinguish between only two states -- a functioning state and a failed state. This dichotomy applies to the structure as well as to each component. The assumption that the state of the system is completely determined by the states of its components implies the existence of a Boolean function $\varphi: B^n\rightarrow B$.

A Boolean function of $n$ variables is a function on $B^n$ taking values in $B=\{0,1\}$. A system or structure is assumed to consist of an element of $\textbf{N}=\{1,2,\ldots,n\}$ - the set of $n$ components. Let us consider the state of the system at a fixed moment of time. 
\begin{definition}
The structure function of a system consisting of $n$ components is a Boolean function of $n$ variables.
\end{definition}

Let $\varphi$ be a structure on $\textbf{N}$ and $i\in\textbf{N}$. The component $i$ is \emph{irrelevant to the structure} $\varphi$ if $\varphi(1,\vecx_{-i})=\varphi(0,\vecx_{-i})$ for all $\vecx\in B^n$ and relevant otherwise. The number of relevant components is called the order of the structure $\varphi$. The structure with no relevant components is a \emph{degenerate structure}, i.e., $\varphi(\vecx)=1$ or  $\varphi(\vecx)=0$ for all $\vecx\in B^n$.

Let $\varphi_i$, $i=1,2$ be two structures on $\textbf{N}=\{1,2,\ldots,n\}$. The linear composition of these two structures is a structure $h(\vecx,x_{n+1})=x_{n+1}\varphi(\vecx)+(1-x_{n+1})\varphi_2(\vecx)$ on $\textbf{B}^{n+1}$.
\begin{corollary}
Any structure $\varphi$ of order $n$ is a linear composition of two structures of at most order $n-1$: 
\begin{equation}\label{decompB}
\varphi(\vecx)=x_i\varphi(1,\vecx_{-i})+(1-x_i)\varphi(0,\vecx_{-i}), \text{for every $\vecx\in B^n$, $i\in\textbf{N}$.}
\end{equation}
\end{corollary}
\begin{definition}
Let $\varphi$ be a structure on $\textbf{N}$, $A\subset\textbf{N}$ and $J = \textbf{N}\setminus A$. The collection of  $A$ of components form a
path (cut) set of $\varphi$ if $\varphi(\vec1^A,\vec0^J) = 1$ ($\varphi(\vec0^A,\vec1^J) = 0$).
\end{definition}
\begin{definition}
Let $\varphi$ be a structure on $\textbf{N}$. Its dual $\varphi^{\mathcal{D}}$ is another structure on
$\textbf{N}$ defined by $\varphi^{\mathcal{D}}(\vecx)=1-\varphi(\vec1-\vecx)$ for every $\vecx\in \textbf{B}^n$.
\end{definition}
\begin{definition}
Let $\varphi$ be a structure on $\textbf{N}$. A path (cut) set $S$ of $\varphi$ is called a minimal path (cut) set of $\varphi$ if $T\subset S$ implies that $T$ is not a path (cut) set of a structure $\varphi$.
\end{definition}
The family $\alpha(\varphi)$ ($\beta(\varphi)$) denotes collection of minimal path (cut) sets of the structure $\varphi$. 
\begin{proposition}\label{PropStrutRepresentation1}
For every semicoherent structure, $\varphi$ on $\textbf{N}$ 
\begin{equation}\label{StrutRepresentation1}
\varphi(\vecx) = 1 - \prod_{S\in\alpha(\varphi)} (1 - \prod_{i\in S}x_i) = \prod_{S\in\beta(\varphi)} (1 - \prod_{i\in S}(1-x_i))\text{ for all $\vecx\in \textbf{B}^n$.}
\end{equation}
\end{proposition}

\begin{remark}[A simple form of $\varphi$]
Expanding either one of the two terms on the right hand side of the expression of Proposition~\ref{PropStrutRepresentation1} (putting $x_i^r=x_i$ for $r\geq 1$) we get a structure function in the form
\begin{equation}\label{SimpleStrutRep}
\varphi(\vecx)=\sum_{T\subseteq\textbf{N}}b_T\prod_{j\in T}x_j \text{for all $\vecx\in \textbf{B}^n$} 
\end{equation}
with $b_T$--some integer constants ($\prod_{j\in T}x_j=1$ for $T=\emptyset$).
\end{remark}
For any structure, there always exist at least one simple form and the simple form of a structure is unique.

\section{\label{AppSimpleGame}The simple game}
In game theory considers the set $N=\{1,2,\ldots,n\}$ of players and the power set $2^N$ of coalitions. A function $\lambda:2^N\rightarrow \{0,1\}$ is called a simple game on $N$ in characteristic function form if 
\begin{enumerate}\itemsep1pt \parskip1pt \parsep1pt
    \item[(1)] $\lambda(\emptyset)=0$;
    \item[(2)] $\lambda(N)=1$;
    \item[(3)] $S\subseteq T\subseteq N$ implies $\lambda(S)\leq \lambda(T)$.
\end{enumerate}
A coalition $S\subset N$ is called winning if $\lambda(S)=1$ and it is called blocking if $\lambda(N\setminus S)=0$. Indeed, the collection of winning (or blocking) coalitions in a simple game satisfies the three properties of the basic structure mentioned at the beginning.

\section{\label{PowerIndex}Power indexes}
In the field of game theory \citeauthor{Sha53:value}~(\citeyear{Sha53:value}) and \citeauthor{Ban1965:voting}~(\citeyear{Ban1965:voting}) consider the role of the players in a cooperative game to provide an idea of division of the gain (v. \citeauthor{Ram1990:Simple}~(\citeyear{Ram1990:Simple}). First, Shapley and Shubik examined $n$-player games what let them formulate a characteristic value applicable to simple games. Hence, originally the Shapley-Shubik index measured a power of players in voting games. Their measure is a natural consequence of the influence of a given voter on the result.

\begin{definition}\label{def:banzhaf}
    The Banzhaf index of the $i$-th player [component] denoted as $\psi_i(g)$ is applicable for a semi-coherent structure $g$ on $N$, and is defined by
    \begin{equation}\label{eq:banzhaf}
        \psi_i(g) =  \frac{\eta_i(g)}{2^{n-1}},
    \end{equation}
    where $i\in N$, $r$ is a size of $\eta_i(g)$, which stands for the aggregated sum of critical path vectors of g, and $\eta_i(g) = \sum_{r=1}^{n} \eta_i(r,g)$. 
\end{definition}
Definition \ref{def:banzhaf} is identical to the structural importance \eqref{eqStrImp12} presented by Birnbaum.

\begin{definition}\label{def:shapleyshubik}
    For a semicoherent structure $g$ the Shapley-Shubik index is denoted as $\phi_i(g)$, given by
    \begin{equation}\label{eq:shapleyshubik}
        \phi_i = \sum_{r=1}^{n} \eta_i(r,g) \cdot \frac{ (n-r)!(r-1)!}{n!},
    \end{equation}
    where $i\in N$, $r$ is a size of $\eta_i(r,g)$, which stands for the number of critical path vectors of g. 
\end{definition}
Definition \ref{def:shapleyshubik} is identical to the structural importance \eqref{eqStrImp6} presented by Barlow and Proschan. 

\begin{fact}
If we compare the expressions \eqref{eq:banzhaf} and \eqref{eq:shapleyshubik}, we can see that the index introduced by Shapley-Shubik has a weight $ (n-r)!(r-1)!/n!$ attached to $\eta_i(r,g)$, meanwhile the Banzhaf index is independent on $r$ and has always weight $1/(2^{n-1})$ attached to $\eta_i(r,g)$. Due to the behavior of $ (n-r)!(r-1)!/n!$ for different $n$, we deduce that only very large or very small critical paths may reach the greatest weight.
\end{fact}

Dubey (1975) derived the Shapley-Shubik index as a logical consequence of certain
axioms. Using another set of axioms, Dubey and Shapley (1979) derived the
Banzhaf index. Straflin (1976) using a probabilistic model, providing a unified
framework for power indices. 


\section*{}


\begin{thebibliography}{43}
\providecommand{\natexlab}[1]{#1}
\providecommand{\url}[1]{\texttt{#1}}
\expandafter\ifx\csname urlstyle\endcsname\relax
  \providecommand{\doi}[1]{doi: #1}\else
  \providecommand{\doi}{doi: \begingroup \urlstyle{rm}\Url}\fi

\bibitem[Abouammoh et~al.(1994)Abouammoh, El-Neweihi, and
  Sethuraman]{AboEl-nSet1994:modules}
A.~M. Abouammoh, E.~El-Neweihi, and J.~Sethuraman.
\newblock The role of a group of modules in the failure of systems.
\newblock \emph{Probability in the Engineering and Informational Sciences},
  8\penalty0 (1):\penalty0 89–101, 1994.
\newblock \doi{10.1017/S0269964800003223}.

\bibitem[Amrutkar and Kamalja(2017)]{AmrKam2017:OverviewIM}
K.~P. Amrutkar and K.~K. Kamalja.
\newblock An overview of various importance measures of reliability system.
\newblock \emph{International Journal of Mathematical, Engineering and
  Management Sciences}, 2\penalty0 (3):\penalty0 150--171, 2017.

\bibitem[{Artzner} et~al.(1999){Artzner}, {Delbaen}, {Eber}, and
  {Heath}]{ArtDelEbeHea1999:Coherent}
P.~{Artzner}, F.~{Delbaen}, J.-M. {Eber}, and D.~{Heath}.
\newblock {Coherent measures of risk.}
\newblock \emph{{Math. Finance}}, 9\penalty0 (3):\penalty0 203--228, 1999.
\newblock ISSN 0960-1627; 1467-9965/e.
\newblock \doi{10.1111/1467-9965.00068}.

\bibitem[Banzhaf(1965)]{Ban1965:voting}
J.~Banzhaf.
\newblock Weighted voting doesn’t work: a game theoretic approach.
\newblock \emph{Rutgers Law Rev.}, 19:\penalty0 317--343, 1965.

\bibitem[Barlow and Proschan(1975)]{BarPro1975:Importance}
R.~E. Barlow and F.~Proschan.
\newblock Importance of system components and fault tree events.
\newblock \emph{Stochastic Processes and their Applications}, 3\penalty0
  (2):\penalty0 153--173, 1975.
\newblock ISSN 0304-4149.
\newblock \doi{10.1016/0304-4149(75)90013-7}.

\bibitem[Barlow and Wu(1978)]{1978CSwM}
R.~E. Barlow and A.~S. Wu.
\newblock Coherent systems with multi-state components.
\newblock \emph{Mathematics of Operations Research}, 3\penalty0 (4):\penalty0
  275--281, 1978.
\newblock ISSN 0364-765X.
\newblock \doi{10.1287/moor.3.4.275}.

\bibitem[Barlow et~al.(1975)Barlow, Fussell, and
  Singpurwalla]{BarFusSin1975:DedBirnbaum}
R.~E. Barlow, J.~B. Fussell, and N.~D. Singpurwalla, editors.
\newblock \emph{Reliability and fault tree analysis}.
\newblock Society for Industrial and Applied Mathematics, Philadelphia, Pa.,
  1975.
\newblock Theoretical and applied aspects of system reliability and safety
  assessment, Conference held at the University of California, Berkeley,
  Calif., Sept. 3--7, 1974, Dedicated to Professor Z. W. Birnbaum.

\bibitem[Bergman(1985)]{Ber1985:NewImportance}
B.~Bergman.
\newblock On some new reliability importance measure.
\newblock In W.~Quirk, editor, \emph{Safety of computer control systems 1985
  (SAFECOMP'85). Proc. of the Fourth IFAC Workshop (Como, Italy, 1-3 October
  1985)}, volume 4: achieving safe real time computer systems, pages 61--64.
  Elsevier, 1985.

\bibitem[{Birnbaum}(1968)]{Bir1968:importance}
Z.~W. {Birnbaum}.
\newblock {On the importance of components in a system.}
\newblock {Eur. Meet. 1968, Sel. Stat. Pap. 2, 83-95 (1968).}, 1968.

\bibitem[Birnbaum(1969)]{Bir1969:Multivariate}
Z.~W. Birnbaum.
\newblock On the importance of different components in a multicomponent system.
\newblock In P.~Krishnaiah, editor, \emph{Multivariate {A}nalysis, {II}
  ({P}roc. {S}econd {I}nternat. {S}ympos., {D}ayton, {O}hio, 1968)}, pages
  581--592. Academic Press, New York, 1969.

\bibitem[Birnbaum et~al.(1961)Birnbaum, Esary, and
  Saunders]{BirnbaumZ.W.1961MSaS}
Z.~W. Birnbaum, J.~D. Esary, and S.~C. Saunders.
\newblock Multi-component systems and structures and their reliability.
\newblock \emph{Technometrics}, 3\penalty0 (1):\penalty0 55--77, 1961.
\newblock ISSN 0040-1706.

\bibitem[{Boland} et~al.(1988){Boland}, {El-Neweihi}, and
  {Proschan}]{BolEl-NPro1988:Active}
P.~J. {Boland}, E.~{El-Neweihi}, and F.~{Proschan}.
\newblock {Active redundancy allocation in coherent systems}.
\newblock \emph{{Probab. Eng. Inf. Sci.}}, 2\penalty0 (3):\penalty0 343--353,
  1988.
\newblock ISSN 0269-9648; 1469-8951/e.

\bibitem[Cao et~al.(2019)Cao, Liu, and Fang]{CaoLiuFan2019:modules}
Y.~Cao, S.~Liu, and Z.~Fang.
\newblock Importance measures for degrading components based on cooperative
  game theory.
\newblock \emph{International Journal of Quality \& Reliability Management},
  37\penalty0 (2):\penalty0 189–206, 2019.
\newblock \doi{10.1108/IJQRM-10-2018-0278}.

\bibitem[Do and Bérenguer(2020)]{DoBer2020:Conditional}
P.~Do and C.~Bérenguer.
\newblock Conditional reliability-based importance measures.
\newblock \emph{Reliability Engineering \& System Safety}, 193:\penalty0
  106633, 2020.
\newblock ISSN 0951-8320.
\newblock \doi{10.1016/j.ress.2019.106633}.

\bibitem[Dresher(1981)]{dresh81:games}
M.~Dresher.
\newblock \emph{The mathematics of games of strategy}.
\newblock Dover Publications, Inc., New York, 1981.
\newblock ISBN 0-486-64216-X.
\newblock Theory and applications, Reprint of the 1961 original. \MR{671740}.

\bibitem[Dui et~al.(2017)Dui, Si, Wu, and Yam]{DuiSiWuYam2017:Importance}
H.~Dui, S.~Si, S.~Wu, and R.~Yam.
\newblock An importance measure for multistate systems with external factors.
\newblock \emph{Reliability Engineering \& System Safety}, 167:\penalty0 49 --
  57, 2017.
\newblock ISSN 0951-8320.
\newblock \doi{10.1016/j.ress.2017.05.016}.
\newblock Special Section: Applications of Probabilistic Graphical Models in
  Dependability, Diagnosis and Prognosis.

\bibitem[Dutuit and Rauzy(2015)]{DutRau2015:ExtensionIM}
Y.~Dutuit and A.~Rauzy.
\newblock On the extension of importance measures to complex components.
\newblock \emph{Reliability Engineering \& System Safety}, 142:\penalty0 161 --
  168, 2015.
\newblock ISSN 0951-8320.
\newblock \doi{10.1016/j.ress.2015.04.016}.

\bibitem[{El-Neweihi} and {Sethuraman}(1991)]{El-NSet1991:modules}
E.~{El-Neweihi} and J.~{Sethuraman}.
\newblock {A study of the role of modules in the failure of systems}.
\newblock \emph{{Probab. Eng. Inf. Sci.}}, 5\penalty0 (2):\penalty0 215--227,
  1991.
\newblock ISSN 0269-9648; 1469-8951/e.

\bibitem[{El-Neweihi} et~al.(1978){El-Neweihi}, {Proschan}, and
  {Sethuraman}]{El-NProSet1978:simple}
E.~{El-Neweihi}, F.~{Proschan}, and J.~{Sethuraman}.
\newblock {A simple model with applications in structural reliability,
  extinction of species, inventory depletion and urn sampling}.
\newblock \emph{{Adv. Appl. Probab.}}, 10:\penalty0 232--254, 1978.
\newblock ISSN 0001-8678.

\bibitem[{Esary} and {Proschan}(1963)]{EsaPro1963:Coherent}
J.~D. {Esary} and F.~{Proschan}.
\newblock {Coherent structures of non-identical components.}
\newblock \emph{{Technometrics}}, 5:\penalty0 191--209, 1963.
\newblock ISSN 0040-1706; 1537-2723/e.

\bibitem[Fussell and Vesely(1972)]{FusVes1972:Overview}
J.~Fussell and W.~Vesely.
\newblock New methodology for obtaining cut sets for fault trees.
\newblock \emph{Trans. Amer. Nucl. Soc.}, 15\penalty0 (1):\penalty0 262--263,
  1972.
\newblock 18. annual American Nuclear Society conference; Las Vegas, Nev; 18
  Jun 1972.

\bibitem[Kordecki(2002)]{kordecki}
W.~Kordecki.
\newblock Oszacowania niezawodności systemów.
\newblock \emph{Mathematica Applicanda}, 30\penalty0 (44/03), 2002.
\newblock \doi{10.14708/ma.v30i44/03.1902}.

\bibitem[{Moulin}(1982)]{Mou1982:GT}
H.~{Moulin}.
\newblock \emph{{Game theory for the social sciences. }}.
\newblock Studies in Game Theory and Mathematical Economics. New York
  University Press, New York, 1982.
\newblock ISBN 0-8147-5386-8/hbk; 0-8147-5387-6/pbk.
\newblock Transl. from the French by the author. \ZBL{0626.90095}.

\bibitem[Nash(1951)]{nas51:noncoop}
J.~Nash.
\newblock Non-cooperative games.
\newblock \emph{Ann. of Math. (2)}, 54:\penalty0 286--295, 1951.
\newblock ISSN 0003-486X.
\newblock \doi{10.2307/1969529}.
\newblock \MR{43432}.

\bibitem[Natvig(1985)]{Nat1985:New}
B.~Natvig.
\newblock New light on measures of importance of system components.
\newblock \emph{Scand. J. Statist.}, 12\penalty0 (1):\penalty0 43--54, 1985.
\newblock ISSN 0303-6898.
\newblock \MR{804224}.

\bibitem[{Navarro} et~al.(2019){Navarro}, {Arriaza}, and
  {Su\'arez-Llorens}]{NavArrSua2019:Minimal}
J.~{Navarro}, A.~{Arriaza}, and A.~{Su\'arez-Llorens}.
\newblock Minimal repair of failed components in coherent systems.
\newblock \emph{Eur. J. Oper. Res.}, 279\penalty0 (3):\penalty0 951--964, 2019.
\newblock ISSN 0377-2217.
\newblock \ZBL{1430.90217}.

\bibitem[Ohi(2010)]{OhiFumio2010}
F.~Ohi.
\newblock Multistate coherent systems.
\newblock In \emph{Stochastic Reliability Modeling, Optimization And
  Applications}, pages 3--34. World Scientific Publishing Co. Pte. Ltd., 2010.
\newblock ISBN 9789814277440.

\bibitem[Owen(2013)]{Owe2013:GT}
G.~Owen.
\newblock \emph{Game theory}.
\newblock Emerald Group Publishing Limited, Bingley, fourth edition, 2013.
\newblock ISBN 987-1-7819-0507-4.
\newblock \MR{3443071}.

\bibitem[Ping(2004)]{Pin2004:PhD}
Z.~Ping.
\newblock
  \emph{\href{http://gateway.proquest.com/openurl?url_ver=Z39.88-2004&rft_val_fmt=info:ofi/fmt:kev:mtx:dissertation&res_dat=xri:pqdiss&rft_dat=xri:pqdiss:3126467}{Measures
  of importance with applications to inspection policies}}.
\newblock ProQuest LLC, Ann Arbor, MI, 2004.
\newblock ISBN 978-0496-73752-9.
\newblock URL \url{https://www.proquest.com/docview/305074799}.
\newblock Thesis (Ph.D.)--University of Illinois at Chicago. \MR{2705807}.

\bibitem[Ramamurthy(1990)]{Ram1990:Simple}
K.~G. Ramamurthy.
\newblock \emph{Coherent structures and simple games}, volume~6 of \emph{Theory
  and Decision Library. Series C: Game Theory, Mathematical Programming and
  Operations Research}.
\newblock Kluwer Academic Publishers Group, Dordrecht, 1990.
\newblock ISBN 0-7923-0869-7.
\newblock \doi{10.1007/978-94-009-2099-6}.

\bibitem[{Rausand} et~al.(2021){Rausand}, {Barros}, and
  {H{\o}yland}]{RauBarHoy2021:book}
M.~{Rausand}, A.~{Barros}, and A.~{H{\o}yland}.
\newblock \emph{{System reliability theory. Models, statistical methods, and
  applications.}}
\newblock Hoboken, NJ: John Wiley \& Sons, 3rd edition edition, 2021.
\newblock ISBN 978-1-119-37352-0.
\newblock 2nd edition: ISBN 0-471-47133-X. xix, 636 p. (2004).

\bibitem[Średnicka(2019)]{Sre2019:MSc}
M.~Średnicka.
\newblock {Importance measure in multi-state systems reliability}.
\newblock Master's thesis, Wrocław University of Science and Technology,
  Wrocław, Poland, 2019.
\newblock 40 p.

\bibitem[Schmidt et~al.(1985)Schmidt, Jamali, Parry, and
  Gibbon]{EPRINP3912SRVol2}
E.~Schmidt, K.~Jamali, G.~Parry, and S.~Gibbon.
\newblock {Importance measures for use in PRAs and risk management}.
\newblock Technical Report~6, Electric Power Research Inst., Palo Alto, CA
  (USA), 1985.
\newblock Proceedings EPRI-NP--3912-SR-Vol2. Sessions 9-16. p. 83.1--83.11.

\bibitem[Shapley(1953)]{Sha53:value}
L.~S. Shapley.
\newblock A value for n--person games.
\newblock In H.~Kuhn and A.~Tucker, editors, \emph{Contrib. Theory of Games},
  Ann. Math. Stud. no 28, pages 307--317. Princeton University Press,
  Princeton, 1953.

\bibitem[{Shiryayev}(1978)]{shi78:book}
A.~N. {Shiryayev}.
\newblock \emph{Optimal Stopping Rules.}
\newblock Springer, New York, 1978.
\newblock English translation of \emph{ \cyr Statisticheski{\u i}
  posledovatelny{\u i} analiz} by A. B. Aries.

\bibitem[Spivak(1965)]{spivak1965calculus}
M.~Spivak.
\newblock \emph{Calculus On Manifolds: A Modern Approach To Classical Theorems
  Of Advanced Calculus}.
\newblock Avalon Publishing, 1965.
\newblock ISBN 9780805390216.
\newblock URL \url{https://books.google.com.vc/books?id=PBcbMQAACAAJ}.

\bibitem[Szajowski and Yasuda(1997)]{szayas95:voting}
K.~Szajowski and M.~Yasuda.
\newblock {V}oting procedure on stopping games of {M}arkov chain.
\newblock In S.~O. Anthony H.~Christer and L.~C. Thomas, editors,
  \emph{UK-Japanese Research Workshop on Stochastic Modelling in Innovative
  Manufacturing, July 21-22, 1995}, volume 445 of \emph{Lecture Notes in
  Economics and Mathematical Systems}, pages 68--80. Moller Centre, Churchill
  College, Univ. Cambridge, UK, Springer, 1997.
\newblock \doi{10.1007/978-3-642-59105-1\_6}.
\newblock \MR{98a:90159}; \ZBL{0878.90112}.

\bibitem[{Szajowski}(2020)]{Sza2020:Rationalization}
K.~J. {Szajowski}.
\newblock Rationalization of detection of the multiple disorders.
\newblock \emph{{Stat. Pap.}}, 61\penalty0 (4):\penalty0 1545--1563, 2020.
\newblock ISSN 0932-5026; 1613-9798/e.
\newblock \doi{https://doi.org/10.1007/s00362-020-01168-2}.
\newblock \ZBL{1448.91017}.

\bibitem[Tijs(2003)]{Tij2003:GTIntro}
S.~Tijs.
\newblock \emph{Introduction to game theory}, volume~23 of \emph{Texts and
  Readings in Mathematics}.
\newblock Hindustan Book Agency, New Delhi, 2003.
\newblock ISBN 81-85931-37-2.

\bibitem[Wu and Coolen(2013)]{WuCoo2013:Cost-based}
S.~Wu and F.~P. Coolen.
\newblock {A cost-based importance measure for system components: An extension
  of the Birnbaum importance}.
\newblock \emph{European Journal of Operational Research}, 225\penalty0
  (1):\penalty0 189 -- 195, 2013.
\newblock ISSN 0377-2217.
\newblock \doi{10.1016/j.ejor.2012.09.034}.

\bibitem[{Xie}(1987)]{Xie1987:Importance}
M.~{Xie}.
\newblock {On some importance measures of system components.}
\newblock \emph{{Stochastic Processes Appl.}}, 25:\penalty0 273--280, 1987.
\newblock ISSN 0304-4149.
\newblock \doi{10.1016/0304-4149(87)90205-5}.

\bibitem[Yasuda et~al.(1982)Yasuda, Nakagami, and
  Kurano]{YasNakKur1982:Multivariate}
M.~Yasuda, J.~Nakagami, and M.~Kurano.
\newblock Multivariate stopping problems with a monotone rule.
\newblock \emph{J. Oper. Res. Soc. Japan}, 25\penalty0 (4):\penalty0 334--350,
  1982.
\newblock ISSN 0453-4514.
\newblock \doi{10.15807/jorsj.25.334}.
\newblock \MR{692543}.

\bibitem[Załęska-Fornal(2006)]{zaleskafornal}
A.~Załęska-Fornal.
\newblock Miary niezawodnościowej i strukturalnej istotności elementów.
\newblock \emph{Zeszyty Naukowe Akademii Marynarki Wojennej}, R. 47\penalty0 (3
  (166)):\penalty0 137--150, 2006.

\end{thebibliography}
\end{document}